\documentclass[12pt,preprint,psfig]{aastex}
\newcommand{\etal}{{\it et al.} }
\newcommand{\rxte}{{\it RXTE} }
\newcommand{\rxtep}{{\it RXTE}}
\newcommand{\asca}{{\it ASCA} }
\newcommand{\ascap}{{\it ASCA}}

\newcommand{\xmm}{{\it XMM-Newton} }

\newcommand{\chandrap}{{\it Chandra}}

\newcommand{\feka}{{Fe~K$\alpha$} }

\newcommand{\bsax}{{\it BeppoSAX} }
\newcommand{\bsaxp}{{\it BeppoSAX}}
\newcommand{\fighistlc}{{Fig.~1} }
\newcommand{\fighistlcp}{{Fig.~1}}
\newcommand{\figgvsg}{{Fig.~2a} }

\newcommand{\figivsi}{{Fig.~2b} }

\newcommand{\figxteltcrv}{{Fig.~3} }
\newcommand{\figxteltcrvp}{{Fig.~3}}

\newcommand{\figgvsfp}{{Fig.~4}}
\newcommand{\figevsl}{{Fig.~5} }
\newcommand{\figevslp}{{Fig.~5}}
\newcommand{\figivsl}{{Fig.~6} }
\newcommand{\figivslp}{{Fig.~6}}

\newcommand{\figewvslp}{{Fig.~8}}
\newcommand{\figresultsvst}{{Fig.~7} }

\newcommand{\figgrpplratp}{{Fig.~9}}
\newcommand{\figgrprat}{{Fig.~10} }
\newcommand{\figgrpratp}{{Fig.~10}}

\newcommand{\figivsecontp}{{Fig.~11}}

\newcommand{\figibvsincontp}{{Fig.~12}}

\newcommand{\figidvsroutp}{{Fig.~13}}

\newcommand{\figfeatewvssig}{{Fig.~14} }

\newcommand{\tableobslogp}{Table~1}
\newcommand{\tableresultslog}{Table~2 }
\newcommand{\tableresultslogp}{Table~2}
\newcommand{\tablegrpresultslog}{Table~3 }
\newcommand{\tablegrpresultslogp}{Table~3}

\newcommand{\sigfeka}{$\sigma_{\rm Fe \ K\alpha}$ }
\newcommand{\sigfekap}{$\sigma_{\rm Fe \ K\alpha}$}
\newcommand{\efeka}{$E_{\rm Fe \ K\alpha}$ }
\newcommand{\efekap}{$E_{\rm Fe \ K\alpha}$}
\newcommand{\ifeka}{$I_{\rm Fe \ K\alpha}$ }
\newcommand{\ifekap}{$I_{\rm Fe \ K\alpha}$}

\begin{document}
\title{Monitoring the Violent Activity from the Inner Accretion Disk of the Seyfert 1.9 Galaxy NGC~2992 with RXTE}
\author{Kendrah D. Murphy\altaffilmark{1}, Tahir Yaqoob\altaffilmark{1,2}, \& Yuichi
Terashima\altaffilmark{3} }
\altaffiltext{1}{Department of Physics and Astronomy,
Johns Hopkins University, Baltimore, MD 21218}
\altaffiltext{2}{Astrophysics Science Division,
NASA/Goddard Space Flight Center, Greenbelt, MD 20771}
\altaffiltext{3}{Department of Physics, Ehime University, Bunkyo-cho, Matsuyama, Ehime 790-8577,
Japan}

\begin{abstract} 

We present the results of a one year monitoring campaign of the Seyfert 1.9 galaxy NGC~2992
with {\it RXTE}.  Historically, the source has been shown to vary dramatically in 2--10 keV
flux over timescales of years and was thought to be slowly transitioning between periods of
quiescence and active accretion.  Our results show that in one year the source continuum flux
covered almost the entire historical range, making it unlikely that the low-luminosity states
correspond to the accretion mechanism switching off.  During flaring episodes we found that a
highly redshifted Fe~K line appears, implying that the violent activity is occurring in the
inner accretion disk, within $\sim100$ gravitational radii of the central black hole.  We also
found that the spectral index of the X-ray continuum remained approximately constant
during the large amplitude variability.  These observations make NGC~2992 well-suited for
future multi-waveband monitoring, as a test-bed for constraining accretion models.

\end{abstract}


\section{INTRODUCTION}
\label{intro}

NGC~2992 is a relatively nearby ($z=0.00771$) Seyfert 1.9 galaxy.  Observations of  NGC~2992
have been made by every X-ray mission since its discovery by $HEAO-1$ in 1977 (Piccinotti
\etal 1982). \fighistlc shows the historical light curve (2--10 keV flux) for X-ray
observations over the past $\sim30$ years (Piccinotti \etal 1982, Mushotzky 1982, Turner \&
Pounds 1989, Turner \etal 1991, Nandra \& Pounds 1994, Weaver \etal 1996, Gilli \etal 2000,
Yaqoob \etal 2007).  In 1994, \asca observed NGC~2992 to have a very low X-ray continuum flux
compared with previous observations (Weaver \etal 1996), down by a factor of $\sim20$ since
it was first detected.  The \asca data revealed a very prominent, narrow \feka emission line,
believed to have originated in matter distant from the supermassive black hole (possibly in
the putative obscuring torus).  Later, X-ray observations with \bsax (Gilli \etal 2000)
showed a revival to a high-flux state.  This was accompanied by complex variability in the
\feka line profile, which appeared to have possible contributions from radiation originating
in both the accretion disk and distant matter.  Gilli \etal (2000) suggested that the
variation in continuum flux was evidence of quenching and revival of accretion on a timescale
of years.  In fact, they believed that \bsax witnessed a revival of the AGN between its
two observations, which were separated by about a year in the period 1997--1998.  Not shown
in \fighistlc is a third observation made by \bsax during its last week of operation in
April 2002, from which only PDS data were taken (see Beckmann \etal 2007). 
Hard X-ray data taken by {\it INTEGRAL} (in 2005 May) and {\it Swift} (in 2005--2006) are also
presented by Beckmann \etal (2007). 

We obtained observing time for a one-year monitoring campaign of NGC~2992 with the Rossi X-ray
Timing Explorer ({\it RXTE}) beginning in 2005.  NGC~2992 had never been observed with \rxte
prior to this campaign.  The goal was to test the response of the \feka line to the variation
in the X-ray continuum.  Spectral analysis of these data revealed an unexpected result: the
variation in flux for the twenty-four \rxte observations covered nearly the entire dynamical
range in flux in the historical data in less than a year.  Previously it was thought that the
variation occurred over decades, but the \rxte data showed flares in the X-ray continuum flux
on timescales of days.  Additionally, the \feka emission line (that is present in the majority
of the observations and centered at $\sim6.4$ keV in most cases) was redshifted and broadened
during high-flux periods, implying that we may have been witnessing short-term flaring
activity (on the order of days to weeks) from the inner regions of the accretion disk.

In \S\ref{obs}, we describe the \rxte data and our data reduction process.  We describe the
results of our spectral analysis of the data in \S\ref{fitting}, including a discussion of
the variability of the \feka emission line and the physical implications.  In
\S\ref{conclusions} we summarize the conclusions of our analysis.  An Appendix is included to
discuss a calibration feature found in the data in the 8--9 keV range.

\section{OBSERVATIONS \& DATA REDUCTION}
\label{obs}

NGC~2992 was observed between 2005 March 4 and 2006 January 28 with the \rxte Proportional
Counter Array (PCA; Jahoda \etal 2006). We obtained data from a total of twenty-four
observations. The observation log is given in \tableobslogp; hereafter we will refer to each
observation as obs~1 to 24 as listed in \tableobslogp.  It can be seen that the interval
between observations ranged from $\sim 3$ to 33 days.   

The PCA is an array of five Proportional Counter Units (PCUs), each with a net geometric
collecting area of $\sim1600$ cm$^{2}$ (Jahoda \etal 2006).  Each PCU consists of 3 layers of a
mixture of Xenon (90\%) and Methane (10\%) gas and a $1^{\circ}$ collimator (FWHM).  For
energies less than 10 keV, $\sim$ 90\% of the cosmic photons and 50\% of the internal
instrumental background are detected in the top layer (layer 1) of the Xenon/Methane detector. 
Therefore, in order to maximize the signal-to-noise ratio, only data from layer 1 of the PCUs
were used.

Fewer of the PCUs were in operation later in the mission, so we did not obtain data from all
five units.  For two observations (obs~2 and obs~3), PCUs 0, 1, and 2 were operational;
however, only  PCU~0 and PCU~2 were operational for {\it all} twenty-four of the NGC~2992
observations.  In order to perform a uniform analysis of all of the observations, we
discarded the PCU~1 data for obs~2 and obs~3, and only used data from PCU~0 and PCU~2.

In the spring of 2000 (May 12), the propane layer in PCU~0 lost pressure and thus the
calibration and background subtraction model were adversely affected (Jahoda \etal 2006). 
Since then, the calibration and background estimation have improved, so we retained the PCU~0
data, but performed initial analyses on PCU~0 and PCU~2 separately, in order to check for
inconsistencies in the two detectors.  We found the variation in the results from spectral
fits to the data from the two detectors to be statistically insignificant.  As we will
describe in \S\ref{fitting}, we analyzed the spectra from the combined PCU~0 plus PCU~2 data
in addition to the individual PCU data in order to obtain better statistics on the important
model  parameters. 

The data were selected to exclude time intervals when the Earth's elevation angle was less than
$10^{\circ}$, during passage of the satellite through the South Atlantic Anomaly (SAA), and
during times of high electron contamination (ie, the housekeeping parameter `ELECTRON2' was
selected to be greater than 0.1).  The background was estimated with the program PCABACKEST,
using the mission-long `Faint' model file pca\_bkgd\_cmfaintl7\_eMv20031123.mdl (Jahoda \etal
2006).  Spectral response matrices were made for each of the observations for both PCU~0 and
PCU~2 using pcarmf version 10.1 and the channel-to-energy calibration file
pca\_e2c\_e05v03.fits.

We extracted spectra and light curves for PCU~0 and PCU~2 for each of the twenty-four
observations.  In addition, we co-added the PCU~0 and PCU~2 data to make twenty-four
combined (PCU~0~+~PCU~2) spectra.  In obs~7 and obs~15, we found flares in the PCU~0 light
curves, so we applied further data selection with time cuts to remove those periods from
the PCU~0 data.  These selection criteria resulted in net exposure times for the
PCU~0~+~PCU~2 data in the range of $\sim1.5$ to $\sim6.5$ ks, as shown in \tableobslogp. 
The grand total integration time for the twenty-four observations, summed over
PCU~0~+~PCU~2, was 133.088 ks.  The mean, full-band, background-subtracted count rates for
the combined PCU~0~+~PCU~2 data varied by over an order  of magnitude, ranging from
$\sim0.8$ to $\sim10.6$ cts/s (see \tableobslogp).  

\section{SPECTRAL FITTING RESULTS}
\label{fitting}
 
We modeled the PCA spectra using XSPEC version 11.3.2 (Arnaud 1996).  Preliminary
examination of the spectra revealed poor background subtraction above $\sim15$ keV.  Since
the PCA data below $\sim3$ keV are not well calibrated (Jahoda \etal 2006), we performed
spectral fitting in the 3--15 keV range.  We used $\chi^{2}$ as the fit statistic.  In all
of the model fitting the Galactic column was fixed density at $N_{\rm H} = 5.26 \times
10^{20} \rm \ cm^{-2}$ (Dickey \& Lockman 1990).  Although the 3--15 keV energy range is
not sensitive to Galactic absorption, we included it for consistency with models in the
literature that were fitted to lower energy data.  All model parameters will be referred to
in the source frame.  

\subsection{Individual Observation Fits} 
\label{indfits} 

We fitted a simple power-law  model to each of the spectra for the twenty-four observations,
initially for the PCU~0 and PCU~2 data separately and subsequently for the combined PCU
data.  This model consisted of two free parameters, namely the photon index ($\Gamma$) and
the power-law normalization.  We found, in most cases, that the plots of the data/model
ratios of the twenty-four combined PCU spectra showed residuals that may correspond to the
\feka emission line known to be present in the source (e.g. Weaver \etal 1996, Gilli \etal
2000).  Each case was confirmed by examining the separate PCU~0 and PCU~2 fits.

We therefore proceeded to fit a power-law plus a Gaussian emission-line component to the
three sets of twenty-four spectra.  The model consisted of five free parameters: the
power-law normalization, $\Gamma$, the centroid energy of the line (\efekap) in keV, the
intrinsic line width (\sigfekap) in keV, and the intensity of the line (\ifekap) in $\rm
photons \ cm^{-2} \ s^{-1}$.  Although for eight of the twenty-four fits the probability of
obtaining values of $\chi^{2}$ as high as those measured from the data by chance was less
than 10\%, we note that our data do not include compensation for systematic errors so the
goodness of fits cannot be assessed by consideration of the values of $\chi^{2}$ alone.  We
will return to this question when discussing fits to grouped data sets (\S\ref{groupfit}). 
In cases where the line energy could not be constrained, the fit was repeated with the Fe
K$\alpha$ energy fixed at the neutral Fe value of 6.4 keV.  In obs~22--24, where the width of
the line could not be constrained (the line was too weak), we fixed the intrinsic Gaussian
line width at \sigfeka = 0.05 keV in order to obtain statistical errors on the line
intensity.  This width is much less than the PCA energy resolution ($\Delta E \sim 1.02$ keV
FWHM at 6 keV, Jahoda \etal 2006).  For seventeen of the twenty-one remaining data sets, the
decrease in $\chi^{2}$ ranged from 6.3 to 22.2 for the addition of three free parameters for
the line, compared to the continuum-only fits.  This corresponds to detections of the line at
a confidence level greater than 90\% ($\Delta\chi^{2}=6.251$ for three parameters).  Three of
the remaining four observations (obs~2, 11, and 12) had detections of the line at greater
than 68\% confidence ($\Delta\chi^{2}=3.506$ for three parameters) and one (obs~15) had a
detection at only marginally less than 68\% confidence ($\Delta\chi^{2}=3.4$).

We compared the values of $\Gamma$, \efekap, \sigfekap, and \ifeka that were obtained from
spectral fits to the PCU~0 data with those obtained from the PCU~2 data and found no
evidence of systematic differences.  For example, \figgvsg and \figivsi show plots of
$\Gamma$(PCU~0) vs. $\Gamma$(PCU~2) and \ifeka (PCU~0) vs. \ifeka (PCU~2) respectively.  Also,
the ratios of the data to the model for PCU~0 and PCU~2 showed no systematic anomalies within
the statistical errors.  Therefore, hereafter we refer only to the results obtained from the
combined PCU~0~+~PCU~2 spectral fits.  

The power-law plus Gaussian fit results for the combined PCU data are shown in
\tableresultslogp.  Statistical errors for each parameter in \tableresultslog are 68\%
confidence for $n$ interesting parameters, where $n$ = 2 or 3 (corresponding to
$\Delta\chi^{2}$ = 2.279 and 3.506 respectively), depending on the number of
parameters that were fixed in order to obtain stable fits during the error analysis.  The
power-law normalization is not included as an interesting parameter.  Quoting the 68\%
confidence ($1\sigma$) errors facilitates statistical analysis of the results.  We also quote
(in brackets, below the best-fit values) 90\% confidence ranges for one interesting parameter
($\Delta\chi^{2}$ = 2.706), for comparison with other results in the literature.  Parameters
that were fixed are labeled with `$f$' in \tableresultslogp.

For twenty-one of the data sets, statistical errors were found for \sigfeka by fixing the line
energy at the best-fit value and statistical errors were found for \efekap, \ifekap, and
$\Gamma$ by fixing the line width at the best-fit value.  Therefore, for these cases, 68\%
confidence errors were found using $\Delta\chi^{2} = 3.506$ ($n=3$) for all interesting
parameters.  The fits became highly unstable if we allowed both \sigfeka and \efeka to be free
during the error analysis.  A stable fit for the line was not obtained for obs~22, 23, and 24
since the line was not detected with sufficient statistical significance in these
observations.  In these cases, it was necessary to fix both the line energy at 6.4 keV and the
line width at 0.05 keV to obtain statistical errors ($n=2$) for the remaining free
parameters (\ifeka and $\Gamma$).   

\subsection{Light Curve}
\label{ltcrv}

In order to facilitate comparison with the literature, we calculated the 2--10 keV continuum
flux and luminosity values by extrapolating the 3--15 keV models down to 2 keV. The 2--10
keV fluxes and luminosities are presented in \tableresultslogp.  All fluxes\footnote[1]{Note:
\rxte absolute fluxes are systematically higher than \ascap, \bsaxp, \chandrap, and \xmm by
$\sim10-20$\% due to the particular spectrum and normalization adopted for the Crab Nebula for
calibration of the PCA (Jahoda \etal 2006).} are observed-frame values, not corrected for
absorption.  All luminosities are rest-frame values, corrected for absorption.   

The $\sim1$ year light curve (\figxteltcrv) ranges from $\sim0.8 \times 10^{-11} \rm \ to
\sim8.9 \times 10^{-11} \ ergs \ cm^{-2} \ s^{-1}$  in 2--10 keV flux.  The 2--10 keV
luminosity ranges from $0.11 \times 10^{43}$ to $\sim1.17 \times 10^{43}$ ergs $\rm s^{-1}$
(assuming $H_{o}=70$ km s$^{-1}$ Mpc$^{-1}$, $\Lambda=0.73$, $\Omega=1$), while the fraction
of the Eddington luminosity ($L/L_{\rm edd}$) ranges from $\sim1.5 \times 10^{-3}$ to $\sim1.8
\times 10^{-2}$, assuming a black hole mass of $5.2 \times 10^{7} M_{\odot}$ (Woo \& Urry
2002).  Historically, $L/L_{\rm edd}$ has ranged from $\sim0.8 \times 10^{-3}$ to $\sim 1.9
\times 10^{-2}$.  The first \rxte observation, which had the highest flux, was followed by two
additional high-flux peaks (obs~4 and obs~8).  The flux then dropped to $\sim1\times 10^{-11}
\rm \ ergs \ cm^{-2} \ s^{-1}$ during obs~9 to obs~12.  This was followed by a slight rise in
flux for obs~13 and obs~14.  The flux measurements in the remaining observations stayed in the
$1-2 \times 10^{-11} \ \rm ergs \ cm^{-2} \ s^{-1}$ range, with the last three appearing to
make an upward turn. 

We quantified the variability in the \rxte light curve by calculating the fractional
variability amplitude, $F_{\rm var}$, comparing this measure of variability with results
obtained from \rxte monitoring of other AGN (e.g. see Vaughan \etal 2003; Markowitz \etal 2003;
Markowitz \& Edelson 2004). The quantity $F_{\rm var}$ is simply the square root of the excess
variance, that is also used to characterize AGN variability, taking into account Poisson noise
(e.g. see Turner \etal 1999). Markowitz \& Edelson (2004) calculated $F_{\rm var}$ uniformly
for a sample of AGN, based on 2--12~keV \rxte PCA count rates, and several monitoring
durations, one of which was 216 days. Using the same energy band for NGC~2992, we obtained 
$F_{\rm var} = (91.5 \pm 0.5)\%, (84.0 \pm 0.5)\%$, and  $(41.6 \pm 0.8)\%$ for all 24
observations, the first 16 observations (207 days duration), and the last 14 observations (211
days duration), respectively. Our observations of NGC~2992 were spaced roughly two weeks apart
except for two pairs of observations for which each pair had a separation of only three days.
The count rates from these pairs of observations were averaged in order to obtain roughly equal
spacing in time intervals before calculating $F_{\rm var}$. Markowitz \& Edelson (2004) found
that the majority of the AGN in their sample had 216-day values of $F_{\rm var}$ lying the
range $\sim (1-40)\%$, and the two sources with the highest values of $F_{\rm var}$ of $\sim
60\%$ were NGC~3227 and NGC~4051 (both of these have  X-ray luminosities that are similar to
the luminosity of NGC~2992). Thus the high value of $F_{\rm var}$ obtained for the first part
of the \rxte NGC~2992 campaign (and indeed for the whole campaign) is exceptionally high.
Moreover, the value of $F_{\rm var}$ obtained for the second part of the campaign (in which
NGC~2992 was much ``quieter'') is still at the high end of the range that was obtained by
Markowitz \& Edelson (2004) for their sample of AGN.

We can compare the timescales probed by our \rxte campaign with the expected break frequency,
$\nu_{b}$, given the black-hole mass, based on a correlation observed from studying samples of
Seyfert galaxies and other black-hole systems (e.g. Papadakis 2004 finds $\nu_{b} \sim
15/[M_{\rm BH}/M_{\odot}]$~Hz; see also McHardy \etal 2007 and references therein).  This break
frequency marks the division where the power spectrum flattens from $\sim \nu^{-2}$ to $\sim
\nu^{-1}$ towards lower frequencies. For NGC~2992 the predicted timescale corresponding to the
break frequency is $\sim 40$ days. On the other hand, if we use the relation between the break
timescale and luminosity that McHardy \etal (2007) empirically from studying a sample of AGN,
we obatined $\sim 1-2$ days. Clearly, more frequent sampling than that achieved in our
monitoring of  NGC~2992 is required in order to derive reliable, direct, quantitative
constraints on the  power-spectrum amplitude and break frequency (or timescale).

\subsection{Continuum \& Fe~K Emission Line}
\label{cont}

The photon index, $\Gamma$, ranged from $\sim1.61 \rm \ to \sim2.25$.  A plot of $\Gamma$
versus 2--10 keV flux is given in \figgvsfp.  It is interesting to note that, despite the
significant variation in flux, $\Gamma$ is consistent with a constant value within the
statistical errors.  The weighted average of $\Gamma$ is $1.732 \pm 0.022$.   This behavior
constrasts with some other Seyfert galaxies in which $\Gamma$ increases with continuum
luminosity. Utilizing the 68\% confidence statistical errors, we show in \figevsl the best-fit
line energies plotted against the 2--10 keV luminosity from each observation, except for the
observations in which \efeka had to be fixed (represented by open circles on the plot).  With
the exception of observations that were made when NGC~2992 was in a high-flux state, the line
energies are consistent with 6.4 keV, confirming the presence of an \feka emission line from
cool matter in the source.  In the high-flux spectra, the centroid energy of the line was found
to be $\sim5.6$ keV.  As will be discussed later, this may correspond to a broad component in
the \feka line complex due to redshifted emission resulting from enhanced illumination of the
line-emitting material in close proximity to the central black hole.

The line width, \sigfekap, was a free parameter in twenty-one of the twenty-four combined
PCU~0 plus PCU~2 fits.  For these observations, the best-fit value for \sigfeka ranged from
0.0 to 1.1 keV, often with large statistical errors.  A plot of \feka line intensity
(\ifeka) vs. 2--10 keV luminosity is given in \figivslp.   Although the 2--10 keV
luminosity varies by a factor of $\sim11$, we see from \figivsl that the line intensity
variability amplitude appears to be less than that of the continuum.  While the 68\%
confidence errors of the line intensity covered a range of $0.0 \ \rm to \ 46.2 \times
10^{-5} \ \rm photons \ cm^{-2} \ s^{-1}$, the weighted mean of \ifeka was $\sim6.6 \pm 1.0
\times 10^{-5} \ \rm photons \ cm^{-2} \ s^{-1}$.  The results imply a constant line
intensity or significant time delays between the response of the line emission to continuum
variations.  For comparison of the emission line and continuum variability, we show in
\figresultsvst the results of $L/L_{\rm Eedd}$, $\Gamma$, \efekap, \ifeka for the
twenty-four observations versus time.  As shown in \figewvslp, there is no clear trend in
the variation of the \feka line EW with 2--10 keV continuum luminosity.  Although there is
larger variation at lower luminosities, the uncertainty is also greater.  At higher
luminosities, the EW appears to tend towards values less than 500 eV.
 
We can compare our measured values of the \feka line intensity with the theoretical
expectation in the limit of a Thomson-thin spherically-symmetric distribution of gas
surrounding a central X-ray source.  In the limit of low redshift, the theoretical value of
\ifeka is 
\begin{eqnarray}
\label{eq:ifeka}
\nonumber I_{\rm Fe~K\alpha} & = & 7.44\times10^{-5}
\ h_{70}^{2}
\left(\frac{L_{2-10}}{10^{43} \ \rm ergs \ s^{-1}}\right) 
\left(\frac{<L>}{L}\right)
\left(\frac{N_{\rm H}}{10^{23} \ \rm cm^{-2}}\right)
\left(\frac{\Delta\Omega}{4\pi}\right)
\left(\frac{\omega_{\rm K}}{0.34}\right)
\left(\frac{A_{\rm Fe}}{4.68\times10^{-5}}\right)
\\ & & \left(\frac{3.4}{\Gamma+1.646}\right)
(7.11^{1.7-\Gamma})
\left(\frac{2-\Gamma}{0.3}\right)
\left(\frac{0.764}{10^{2-\Gamma}-2^{2-\Gamma}}\right)
\left(\frac{z}{0.00771}\right)^{-2}
\rm photons \ cm^{-2} \ s^{-1} \hspace{1.5cm} 
\end{eqnarray}
(see for example Yaqoob \etal 2001), where $L_{2-10}$ is the 2--10 keV luminosity  and
$(\Delta\Omega/4\pi)$ is the fraction of the sky covered by the line-emitting region, as
seen by the source.  Here $L$ is the measured luminosity and $<L>$ is representative of the
historically averaged luminosity on timescales greater than the light-crossing time of the
line-emitting region.  The factor $h_{70}$ is defined as $\left(\frac{H_{0}}{70 \ \rm km \
s^{-1} \ Mpc^{-1}}\right)$, where $H_{0}$ is the Hubble constant.  $N_{\rm H}$ is the
column density of the shell.  The value of $A_{\rm Fe}$ is the iron abundance in the emitting
material (relative to hydrogen) and $\omega_{\rm K}$ is the fluorescence yield.  All of the
factors in parentheses involving the photon index in equation \ref{eq:ifeka} evaluate to
$\sim1$ for $\Gamma=1.7$, the approximate mean value found for NGC~2992 from the \rxte
observations. Note that equation~(1) does not represent any assumptions
about NGC~2992. It is simply a theoretical limit with which results from NGC~2992
can be compared. From this comparison one can deduce constraints on the
physical parameters that must apply to NGC~2992.

Overlaid on \figivsl are four theoretical \ifeka versus luminosity curves corresponding to
values of $\left(\frac{<L>}{L}\right)\left(\frac{N_{\rm H}}{10^{23} \ \rm cm^{-2}}\right)$
equal to $0.1,\ 1,\ 5$ and $10$.  The curves were calculated using equation \ref{eq:ifeka} with
$\frac{\Delta\Omega}{4\pi}=1$ and $\Gamma=1.732$, the weighted mean value of the photon
index.  We took $A_{\rm Fe}$ to be $4.68\times10^{-5}$, the solar value given in Anders \&
Grevesse (1989), and we used the value $\omega_{\rm K}=0.34$ as given in, for example, Palmeri
\etal (2003).  The typical line-of-sight $N_{\rm H}$ that has been measured by Suzaku and
other missions is $\sim10^{22} \ \rm cm^{-2}$.  For this value, $\frac{<L>}{L}=1$ corresponds
to the shallowest theoretical line shown on \figivsl (labeled with `0.1'), which is not a good
representation of these data.  Better agreement is given by a larger value of 
$\left(\frac{<L>}{L}\right)\left(\frac{N_{\rm H}}{10^{23} \ \rm cm^{-2}}\right)$, implying
that the line-of-sight $N_{\rm H}$ is not representative of the entire system and/or the
observed variation in the line intensity is affected by time delays following continuum
variations.  Values of $\left(\frac{<L>}{L}\right)\left(\frac{N_{\rm H}}{10^{23} \ \rm
cm^{-2}}\right) \sim 5-10$ are more likely. Weaver \etal (1996) deduced time delays 
of the order of years between the X-ray continuum and distant-matter Fe~K emission line.

For the other extreme, a Compton-thick reprocessor, the Fe~K line equivalent width (EW)
depends on the reprocessor geometry and its orientation with respect to the observer. However,
we can identify two very general scenarios regardless of the details of the geometry. The two
cases correspond to whether or not the structure's orientation and/or geometry is such that it
intercepts the line-of-sight between the X-ray continuum source (that illuminates the
reprocessor to produce the Fe~K emission line) and the observer. If the X-ray continuum is
obscured, then the Fe~K line EW can be in the range of hundreds to thousands of eV, but if it
is not (in which case the Fe~K line is observed in `reflection' - i.e. from the same surface
that is illuminated by the continuum) the EW is typically not more than $\sim 200$~eV. 
Examples of Monte Carlo simulations of such scenarios can be found in Ghisellini \etal (1994)
and references therein. What is apparent from these calculations is that for a given geometry
and orientation, the EW of the Fe~K line, as a function of the value of the highest column
density through the reprocessor, reaches an asymptotic value once the structure becomes
Compton-thick. Therefore, for the purpose of simple estimates and comparison with the
Compton-thin case, one can parameterize all of the Compton-thick models by the asymptotic
value of the Fe~K line EW {\it for the time-steady, static limit for a constant X-ray
continuum}. We can then express the intensity of the Fe~K line in terms of the 2--10~keV
luminosity of a steady-state illuminating X-ray continuum and directly compare that with the
Compton-thin scenario. For the Compton-thick case we get
\begin{eqnarray}
\label{eq:ithick}
\nonumber I_{\rm Fe~K\alpha} & = & 8.00\times10^{-5}
\ h_{70}^{2}
\left(\frac{L_{2-10}}{10^{43} \ \rm ergs \ s^{-1}}\right)
\left(\frac{<L>}{L}\right)
\left(\frac{z}{0.00771}\right)^{-2}
\left(\frac{2-\Gamma}{0.3}\right)
 \\ & &
\left(\frac{0.764}{10^{2-\Gamma}-2^{2-\Gamma}}\right)
\left(\frac{6.4 \ \rm keV}{E_{\rm 0}}\right)^{\Gamma}
\left(6.4\right)^{1.7-\Gamma}
\left(\frac{\rm EW}{100 \ \rm eV}\right)
\rm photons \ cm^{-2} \ s^{-1}
\end{eqnarray}
where $E_{\rm 0}$ is the rest frame energy of the Fe~K line photons in keV.  All of the
factors in parentheses involving the photon index in equation \ref{eq:ithick} evaluate to
$\sim 1$ for $\Gamma=1.7$.

Lines of \feka line intensity versus 2--10~keV luminosity are overlaid on \figivsl for several
values of the asymptotic EW.  This asymptotic EW parameter bundles the unknown information
about the geometry and orientation of the reprocessor into a single number and is degenerate
with the factor $\frac{<L>}{L}$ that represents the ratio between the the historical
luminosity, averaged on timescales longer than the reprocessor light-crossing time, and the
observed continuum luminosity.  We used the weighted mean value of $\Gamma=1.732$ and assumed
that the line arises from neutral Fe. Using equation \ref{eq:ithick}, we calculated
theoretical curves for $\left(\frac{<L>}{L}\right)$EW=100, 250, 500, 750, and 1000 eV.  It can
be seen in \figivsl that the \rxte data points cover the range of theoretical values of the EW
from 100 to 1000 eV for $\frac{<L>}{L}=1$, and therefore the data do not rule out either
extreme of reflection or transmission through a Compton-thick reprocessor.  However, the case
of transmission (EW=1000 eV) appears to be unlikely since this would require the historically
averaged luminosity ($<L>$) to be lower than than the \rxte luminosities in order to fit the
data.  Since the \rxte data covers nearly the entire historical range in luminosity, we would
expect $<L>$ to be larger than the lowest \rxte luminosity values.

The narrow \feka line found by Suzaku had an intensity of $2.49^{+0.71}_{-0.40}
\times10^{-5} \ \rm photons \ cm^{-2} \ s^{-1}$ (Yaqoob \etal 2007), which is generally
smaller than the values obtained by \rxte.  The excess measured by \rxte could be due to part
of the flux from an underlying broad line, which cannot be decoupled from the narrow line due
to the poor PCA resolution.

\subsection{Grouped Data} 
\label{groupfit} 

In order to better constrain the key model parameters and investigate variability in more
detail, we combined groups of the twenty-four spectra into six data sets.  The groups are
identified on \figxteltcrvp.  Group 1 (filled circles) consists of the three high-flux
observations that were completed in the earlier part of 2005 (obs~1, 4, 8).   The
observations in an intermediate-flux state were combined in group 2 (stars; obs~2, 3, 6, 7,
13, 14).   The remaining, low-flux observations were broken up chronologically to create
group 3 (squares; obs~5,  9, 10, 11, 12), group 4 (triangles; obs~15, 16, 17), group 5
(diamonds; obs~18, 19, 20, 21), and  group 6 (open circles; obs~22, 23, 24).  As before, we
fitted a simple power-law to the $3-15$ keV data.  Plots of the ratios of the model to the
grouped data sets are given in \figgrpplratp.  We used different y-axis scales on these
plots in order to clearly display the spectral features.  We found, again, clear residuals
at $\sim6.4$ keV for each of groups 2, 3, 4, and 5.  In group 1 the residuals appeared to
peak at an energy lower than 6.4 keV, while in group 6 no line-like residuals were
evident.  We fitted the data again, adding a Gaussian component to model \feka line
emission.  \tablegrpresultslog gives the results of these fits and the plots of the ratios
of the data to this model are shown in \figgrpratp.  The results were consistent with those
from the individual observations.  The addition of a Gaussian component to the model gave a
better fit for groups 1 through 5, with $\chi^{2}$  decreasing by at least 19.9 for the
addition of three free parameters.  However, the absolute $\chi^{2}$ values remained high. 
Although \tablegrpresultslog shows that the probability of obtaining these values by chance
is very small, the $\chi^{2}$ values do not take into account systematic errors in the data
and therefore are not an adequate assessment of the goodness of the model fits.  The \feka
line was not significantly detected in group 6.  However, we were able to derive a line
intensity with a non-zero lower limit (at 68\% confidence) by fixing the line centroid
energy and width (\sigfekap) at 6.4 keV and 0.05 keV respectively (see
\tablegrpresultslogp).

The 2--10 keV flux ranged from $\sim1.1 \times 10^{-11}$ to $\sim 7.4 \times 10^{-11} \rm \
ergs \ cm^{-2} \ s^{-1}$.  The photon index ranged from $\sim 1.71$ to $\sim 1.96$, with a
weighted mean of $\sim1.733 \pm 0.022$.  The emission line energy appeared to be consistent
with the neutral Fe value of 6.4 keV except for group 1, the high-flux group, where the
centroid peaked at $\sim 5.6$ keV.  Excluding group 6, the best-fit value of \sigfeka
ranged from 0.00 to 0.89 and the best-fit EW of the line ranged from $\sim200$ to $\sim
700$ eV.  The weighted mean value of \ifeka for all six groups was $\sim(6.6 \pm 1.0)
\times 10^{-5} \rm \ photons \ cm^{-2} \ s^{-1}$ and the weighted mean value for groups 2
through 5, where the 6.4 keV Fe~K line dominated, was $\sim(7.5 \pm 1.2) \times 10^{-5}
\rm \ photons \ cm^{-2} \ s^{-1}$.

We directly compared the results from the grouped data fits with the results from the
individual-observation fits by over-plotting the group values on the figures discussed in
\S\ref{indfits}.  The grouped data points are marked with crosses in \figgvsfp, \figevslp,
\figivslp, and \figewvslp.  Like $\Gamma$, the EW of the \feka line did not appear to vary
significantly between the groups (corresponding to a weighted mean value of $310\pm48$ eV
for groups 1 through 5), although the EW values are consistent with the trend of being
smaller for higher continuum luminosities.  However, within the 68\% confidence errors, the
intensity of the \feka line appears to be higher during the high-luminosity observations,
as shown in \figivslp.  In addition, \figevsl shows the distinct difference in the energy
of the \feka line between the high-luminosity grouped data and the lower-luminosity grouped
data.  We investigate this further in \S\ref{highlow}.  

In \figgrprat we see that there sometimes appears to be a statistically significant `dip'
in the spectrum between $\sim 8$ and 9 keV. Even though the residuals in this region are
sometimes as large as those in the Fe K band (when compared to a simple power-law), we
believe that the 8--9 keV dip is an artifact of the background subtraction model and we
are confident that the Fe~K emission-line measurements are reliable because they are in
large part consistent with our knowledge from observations and measurements with other
missions. Our reasons for believing that the 8--9 keV dip is an artifact are detailed in
the Appendix. Since such an artifact in the spectrum can adversely affect fitted values of
the model parameters when a Compton-reflection continuum is included in the model, we
omitted three spectral channels (in the range of 7.7--9.5 keV) in subsequent spectral
fitting to remove the dip.

\subsubsection{Compton Reflection}
\label{reflection}

The effect of Compton reflection is important to consider when modeling spectra of AGNs (e.g.
Reynolds \& Nowak 2003 and references therein).  Although the \rxte spectral band used here is
not very sensitive to a Compton-reflection continuum, we added a reflection component to our
model (hrefl in XSPEC; see Dov{\v c}iak, Karas, \& Yaqoob 2004) in order to ensure that our
results are robust.  The resulting model has six free parameters including the so-called
reflection fraction, $R$, in addition to the power-law normalization, $\Gamma$, \efekap,
\sigfekap, and \ifekap.  $R$ is the normalization of the reflection continuum relative to that
expected from a steady-state, X-ray illuminated, neutral disk subtending a solid angle of
$2\pi$ at the X-ray source.  The Compton-reflection model used here assumes a centrally
illuminated, neutral, infinite disk with the abundance of Fe fixed at the solar value from
Anders \& Grevesse (1989).  We investigated the variation in \ifeka for values of $R$ between 0
to 3 by creating confidence contours of \ifeka versus $R$.  We found in general that the best
fit value of $R$ was consistent with 0 and that it did not significantly impact the value of
\ifekap.  The 99\% confidence contours were open and flat over the range in $R$ considered but
the spectral ratios to a simple power-law model indicated we should have obtained an upper
limit in the searched range of $0 \le R \le 3$, or at least that the contours should have begun
to close. Unfortunately, what is happening is that the limited energy resolution, bandpass and
signal-to-noise ratio of the data results in steep intrinsic continuua with strong reflection
degenerate with intrinsically flatter continuua with little reflection. Thus, upper limits on
$R$ for these data are not necessarily physically meaningful. However, using \bsax data with
PDS coverage out to beyond 100~keV, Gilli \etal (2000) found that Compton reflection was in
fact weak in NGC~2992 (see also Beckmann \etal 2007).  Our main concern here is that the Fe~K
emission line intensities are not sensitive to assumptions about the Compton reflection
continuum and with the above considerations in mind, we did not include Compton reflection in
most of the spectral fits described below (and we will note those for which it is included).
When we do include a Compton reflection continuum in the model, it is only a single component,
although two components might be expected. One component might be associated with the Fe~K line
from an accretion disk, and the other from distant matter (if it is Compton-thick). However the
quality of the data and the restricted bandpass does not warrant modeling multiple
Compton-reflection components so the value of $R$ then represents the relative amplitude of all
reflection components that might be present in the data.

\subsubsection{Variability of the Broad Component of the Fe~K Line}
\label{highlow}

In order to explore the apparent difference between the high- and low-flux states further,
we combined all of the spectra from the observations which made up groups 3, 4, and 5
(low-flux) to compare with the group 1 (high-flux) spectrum. We did not include group 6 in
the low-flux spectrum since the \feka line was not significantly detected in these
observations.  

Confidence contour plots of \ifeka versus \efeka for the high- and low-flux groups are
shown in \figivsecontp.  The model used included Galactic absorption, a power-law
continuum, and a Gaussian line.  Interestingly, the contours for the two sets of data are
mutually exclusive at 99\% confidence.  As can be seen in \figivslp, we did not detect the
same order of magnitude increase in intensity of the \feka line as in the 2--10 keV
continuum luminosity.  We do, however, see a broadening as well as a significant shift in
the centroid energy of the detected \feka line from $\sim6.3$ keV in the low state to
$\sim5.6$ keV in the high state.  It is possible that we are seeing two different
components of the \feka line complex: a component (with a centroid energy near 5.6 keV),
originating in the accretion disk close to the black hole, that is broadened and
redshifted due to Doppler and gravitational effects, and a narrow component (with a
centroid energy closer to 6.4 keV) from more distant material.  The increase in the
redshifted line flux in the high state may be due to a localized flare close to the black
hole.  

We investigated the extent to which the redshifted, broad Fe~K emission component and the
narrow, distant-matter Fe~K line may be present in both the low- and high-flux states by
fitting both high- and low-flux groups with a second Gaussian added to the previous power law
plus Gaussian line model.  We fixed the centroid energy of one Gaussian at 6.3 keV, the
best-fit value for the low-flux group in the previous, single Gaussian fit and we fixed the
other at 5.6 keV, the corresponding best-fit value for the high-flux group.  The confidence
contours of the dual Gaussian fit for the two groups are shown in \figibvsincontp.  The
contours for the two groups are mutually exclusive only up to 90\% confidence.  The contours
show that it is not required that both components exist in both of the data sets (only a broad
line is definitively detected in the high-flux group, only a narrow line in the low-flux
group).  On the other hand, at 99\% confidence, both of the components may be present in both
the high- and the low-flux states. 

In order to determine the implications for the proximity of the flaring region to the black
hole, we fitted a disk-line component in XSPEC (see Fabian \etal~1989) to the high flux data in
order to model the redshifted line.  The parameters of the disk-line model are the rest-frame
line energy ($E_{0}$), the power law index of the emissivity ($q$, where the emissivity$\sim
r^{q}$), the inner and outer radii of the disk emission ($R_{\rm in}$ and $R_{\rm out}$
respectively) relative to the central black hole, and the inclination angle of the disk with
respect to the line-of-sight of the observer ($\theta_{obs}$).  We included a power-law
continuum component in the model as well as a Gaussian component with the line energy (\efekap)
fixed at 6.4 keV and \sigfeka at 0.05 keV in order to model a possible narrow component to the
Fe~K line complex (although none was detected).  For this fit, we also included Compton
reflection (as described in \S\ref{reflection}) since the parameter constraints may be
different when the emission line is modeled as a disk line rather than a Gaussian.  Therefore,
this model had five free parameters: $\Gamma$, \ifekap, the reflection fraction ($R$),
$\theta_{\rm obs}$, and $R_{\rm out}$.  We fixed $E_{0}=6.4$ keV in the rest frame, $R_{\rm
in}=6R_{\rm G}=6GM/c^{2}$, and $q=-1.5$.  This value of $q$ is consistent with the
value found from Suzaku data (see Yaqoob \etal 2007). 
The data certainly do not rule out steeper emissivity laws ($q \sim -3$ to $-2$
may be more typical in AGN), and indeed some theoretical
models predict very steep emissivity laws with $q \sim -6$. 
However, 
$q$ is degenerate with $R_{\rm out}$: steeper values of $q$ will result in
larger allowed values of $R_{\rm out}$ for the same data because steep
emissivity laws by definition give most of the line emission from the innermost
regions of the disk and  $R_{\rm out}$ eventually becomes irrelevant. 
Therefore, by using $q=-1.5$ we are addressing the question of
what is the largest size of the region that can account for the redshifted
Fe~K line because more negative values of $q$ will automatically produce
more of the line from a smaller region for a given value of $R_{\rm out}$.
We found that $\theta_{obs}$ was constrained to be $<39^{\circ}$ (90\% confidence)
and this is not sensitive to $q$ as the same upper limit was obtained
when $q$ was fixed at $-3$.
The inclination angle constraint is consistent with $\theta_{obs}>31^{\circ}$ obtained from
modeling the persistent disk line emission in Suzaku data (Yaqoob \etal 2007).

We plotted confidence contours of the
intensity of the disk line versus the outer radius and found that at 99\% confidence, the line
originated within $\sim100$ gravitational radii of the black hole (see \figidvsroutp). 
Thus, in the context of our interpretation 
that the redshifted broad
component of the Fe~K emission line is due
to the line emission from the inner disk 
in the high state being temporarily enhanced during
continuum flares, we deduce that 
the enhanced region is smaller than $\sim 100$ gravitational radii. 
A similar event was observed in the Seyfert galaxy MCG~-6-30-15 by Iwasawa \etal
(1999) in which the Fe~K line centroid shifted to $\sim5$ keV and the 6.4 keV component was
not the prominent peak as it usually is in that source.  Although the narrow line emission may
have been present during the high-flux state of NGC~2992, it is possible that the continuum
swamped the line so that we were unable to detect it.

The quasi-simultaneous Suzaku observations of NGC~2992 (see \figxteltcrvp; Yaqoob \etal 2007)
showed evidence of a persistent, broad, disk line component with an intensity of
$1.9^{+0.5}_{-1.0}\times10^{-5}$ photons cm$^{-2}$ s$^{-1}$ (90\% confidence for one
interesting parameter).  Such a disk line likely originates from a larger region of the disk
than that found in the \rxte flare spectra and so does not have a highly redshifted centroid
energy compared to 6.4 keV because there is relatively more line emission from larger radii. 
In the \rxte data, we found the upper limits on the intensity of a persistent, extended
($R_{\rm out}=1000 R_{\rm G}$) disk line (at 6.4 keV in the rest frame of the disk) for the
high- and low-flux groups to be $5.4\times10^{-5}$ and $5.8\times10^{-5}$ photons cm$^{-2}$
s$^{-1}$ (90\% confidence for one interesting parameter) respectively.  Thus, the continuous
presence of a persistent disk line with the line intensity observed in the Suzaku data, in
addition to the flaring disk line and the distant matter component, is not ruled out by the
\rxte data.

The flaring behavior in the \rxte data strongly supports the idea that X-rays originate in a
corona above the accretion disk.  If the X-ray source is in fact a corona above the disk, a
flare from the inner region of the disk could significantly intensify a particular part of the
line profile, making the whole line appear to be redshifted.  On the other hand, a flaring,
centrally located source would simply intensify the entire line and therefore the centroid
would not be redshifted. The slope of the power-law continuum ($\Gamma$) 
potentially holds important
information about the temperature ($kT$) and optical depth ($\tau$)
of this plasma from which the X-rays are
thought to originate. 
The spectral index, $\Gamma$, is
proportional to the Compton $y$ parameter which is a function of 
the product $(kT)\tau$.
Inclusion of X-ray data at higher energies than our \rxte data is needed 
because $kT$ could then be constrained directly by the data and then
$\tau$ could be derived from fitting Comptonization models (e.g. see Titarchuk 1994). 
Although high-energy data is available for the two \bsax observations of NGC~2992, Gilli
\etal (2000) did not address the question of constraining the high-energy cut-off of the
power-law continuum (the cut-off was arbitrarily fixed in their model fitting).  However,
Beckmann \etal (2007) find no evidence for a cut-off out to $\sim200$ keV, based on analyses
of \bsax, {\it INTEGRAL}, and {\it Swift} data, implying that the scattering corona is
Compton-thin.  If $\Gamma$ is truly constant during the large amplitude variation (see
\figgvsfp), then this would imply that the Compton $y$ parameter remains steady and any
accretion model must account for this.  

\section{CONCLUSIONS}
\label{conclusions}
\begin{itemize}

\item The \rxte light curve (\figxteltcrvp) shows that the flux varied by approximately a
factor of 10 on timescales of days to weeks. It was previously thought that this AGN went
through an extended period of quiescence over many years, followed by a `rebuilding' of the
accretion disk (see \S\ref{intro}).  However, in less than a year the \rxte data covered nearly
the entire dynamic range in flux seen in the historical data (\fighistlcp), suggesting
short-term flaring activity as opposed to long-term changes in accretion activity. 
Historically, when NGC~2992 was observed in a high-flux state, it exhibited properties of a
type 1 AGN, such as broad optical lines and rapid X-ray variability, as opposed to when it was
in a low-flux state during which it showed properties more consistent with a type 2 AGN.  This
transitioning behavior challenges the unified model and its theoretical predictions.

\item In most of the observations, an emission line was detected at $\sim6.4$ keV, likely
dominated by the \feka emission line that is known to come from distant matter.  The flux
variability of the Fe line was much less than the continuum variability.  Due to the poor
energy resolution of \rxte, the measured intensity of the dominant line component likely
has contributions from matter closer to the black hole (e.g., the accretion disk) as
well.  The best-fitting EW ranged from $\sim200$ to $\sim1200$ eV, which is consistent
with the bulk of the line emission at 6.4 keV having been unresponsive to continuum
variability.

\item In the three highest luminosity observations, a highly redshifted (\efeka $\sim5.6$
keV), broadened Fe line dominated the spectrum and the 6.4 keV component was not detected. 
The redshifted line may be a signature of flaring activity from the inner disk, likely coming
from within $100$ gravitational radii of the black hole (see \figidvsroutp), where strong
gravitational and Doppler effects are important.  Although a distant-matter Fe~K line
component is not detected during the flares, it is not ruled out.  While the high-luminosity
observations may be dominated by flaring in the inner disk, both the low- and high-luminosity
data allow for additional, persistent Fe~K line disk emission from the whole disk, as seen by
Suzaku (Yaqoob \etal 2007).  The \rxte data are not sensitive to this persistent component.

\item Although the continuum luminosity varied by a factor of $\sim 11$, the slope of the
power-law fit was consistent with no variability (weighted mean $\Gamma\sim1.7$).  In terms of
Comptonization models, this implies a roughly constant Compton $y$ parameter, a fact which must be
explained by any general model of accretion onto a supermassive black hole.  

\end{itemize}

Future monitoring of NGC~2992 with higher spectral resolution will be able to resolve the
Fe~K line complex and will improve our understanding of the accretion disk and the structure
in general of this source.  Observations at higher energies (which will allow us to determine
constraints on the plasma temperature and therefore the Compton $y$ parameter, as
well as better constraints on Compton reflection), combined with
multi-waveband monitoring will be critical to constraining black hole accretion models.  

The authors thank Alex Markowitz for useful discussions and help with 
the 3C 273 and 3C 279 data analysis and help with characterizing the
variability of NGC~2992. 
The authors acknowledge partial support from NASA grants NNG05GM34G 
(T.Y., K.M.), NNG0GB78A (T.Y., K.M.), and NRA-00-01-LTSA-034 (T.Y.);
Y.T. is supported by Grants-in-Aid for Scientific
Research (17740121).

\section*{APPENDIX}
\label{append}

Most of the \rxte data sets of NGC~2992 show a dip in both the PCU~0 and PCU~2 spectra near
8--9 keV (see \S\ref{groupfit}).  The strength of this apparent absorption feature relative to
the continuum appears to be variable, with the center depth reaching 0--30\% below the
continuum (see \figgrpplratp). Although there does not appear to be a simple relation between
source flux and dip strength, the dip is always weak ($\sim10$\%) in the high-flux spectra and
is stronger in most of the low-flux spectra.  However, group 6 (see \figxteltcrvp), which was a
low-flux observation where the Fe~K line was not detected, does not show the dip.   

We believe that the dip may be an artifact of the background subtraction model in \rxtep.  The
main evidence for this comes from the Suzaku observations of NGC~2992, which were
quasi-simultaneous with the \rxte group 5 observations (see \figxteltcrvp).  Group 5 is one of
the groups with the strongest dip, yet the Suzaku data did not detect the feature. 
\figfeatewvssig shows the confidence contours of the EW of the feature (when fitted with an
inverse-Gaussian model) versus its intrinsic width for the \rxte group 5 data along with the
99\% confidence upper limit constraints on the Suzaku data (dotted line).  Since the 99\%
confidence contours for the \rxte and Suzaku data do not overlap, the presence of the dip in
the Suzaku data is ruled out.  We found further evidence that the dip is an artifact by
examining some observations of 3C~273 and 3C~279 made in 2005.  We found the same feature in
many of the spectra of these sources, but its strength ranged from 0\% to only roughly 10\% of
the continuum.  Note that attempts to fit the dip feature with astrophysical models (such as
absorption edges) failed to account for the residuals. 

As far as we know, such a dip in the 8--9 keV continuum has not been reported for other
\rxte sources in the literature.  For example, Rothschild \etal (2006) presented \rxte data
for Cen~A from observations made much earlier than those of NGC~2992 but they did not detect
the dip feature.  Therefore, if the dip is an artifact in the \rxte data, it appears to be an
inadequacy of one background model for more recent data. 

\newpage

\begin{figure}[!htb]
 \centerline{
\epsscale{0.5}
\scalebox{0.8}{\includegraphics[angle=270]{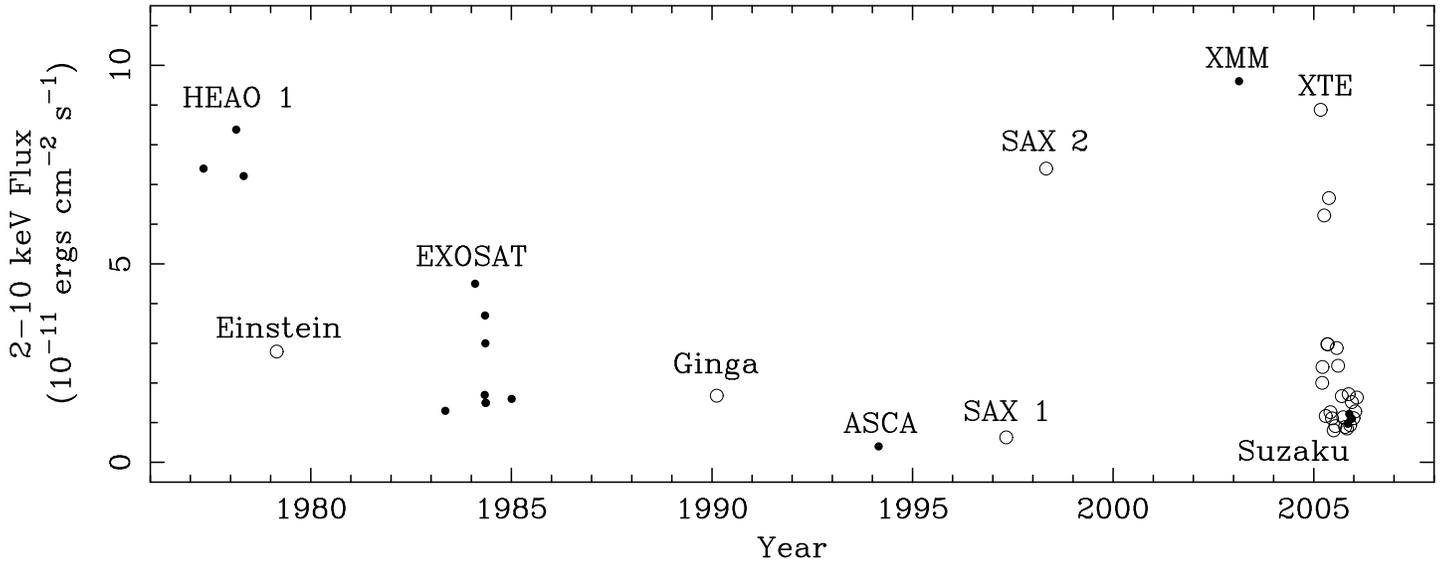}}}
 \caption{
       Light curve of the historical NGC~2992 observations, including the present \rxte campaign.  
       The 2--10 keV flux varies by a factor of $\sim20$ during this $\sim30$ year period.
       }
\end{figure}

\begin{figure}[!htb]
 \centerline{
\epsscale{0.8}
    \plotone{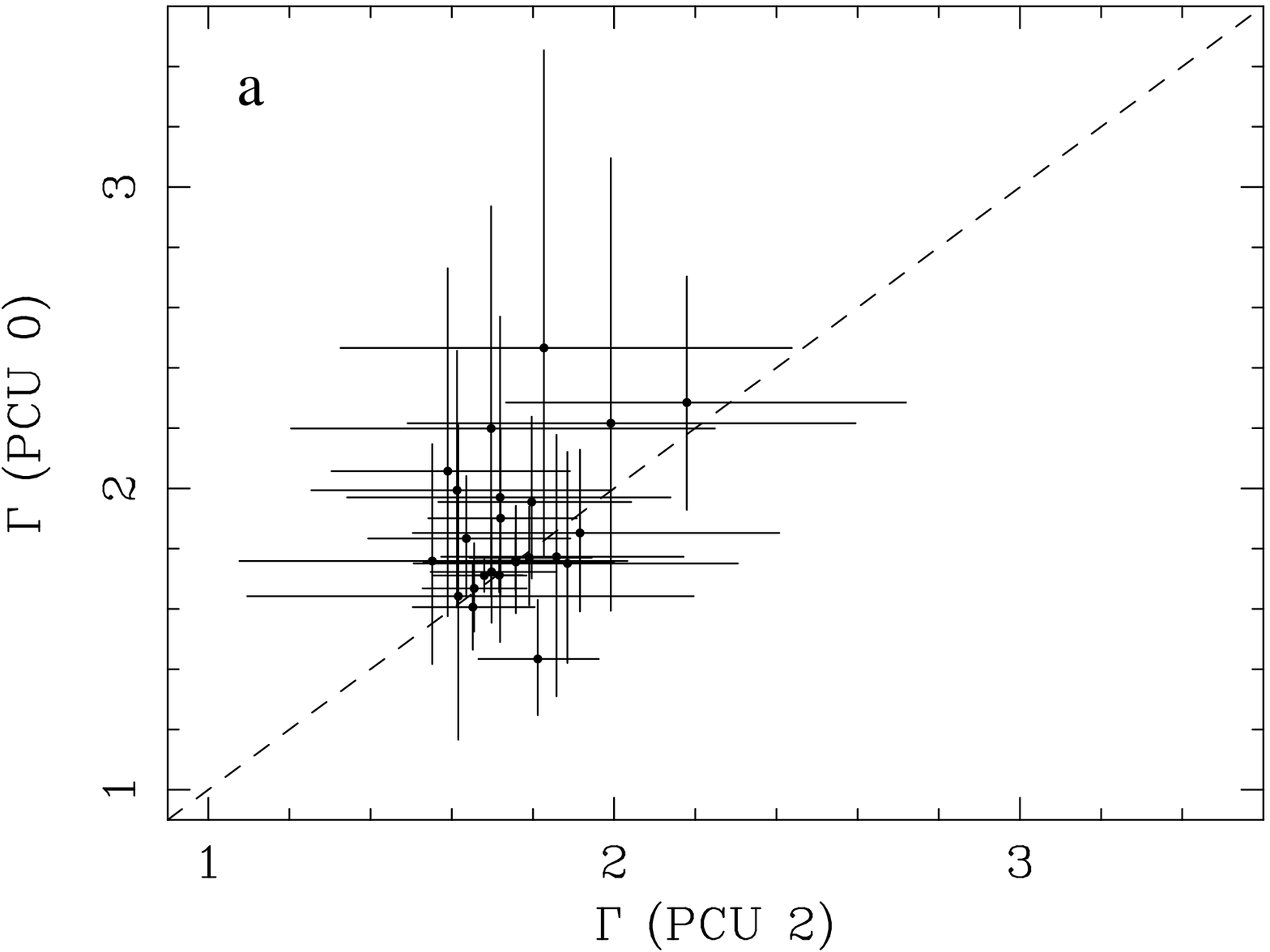}}
 \centerline{
\epsscale{0.8}
    \plotone{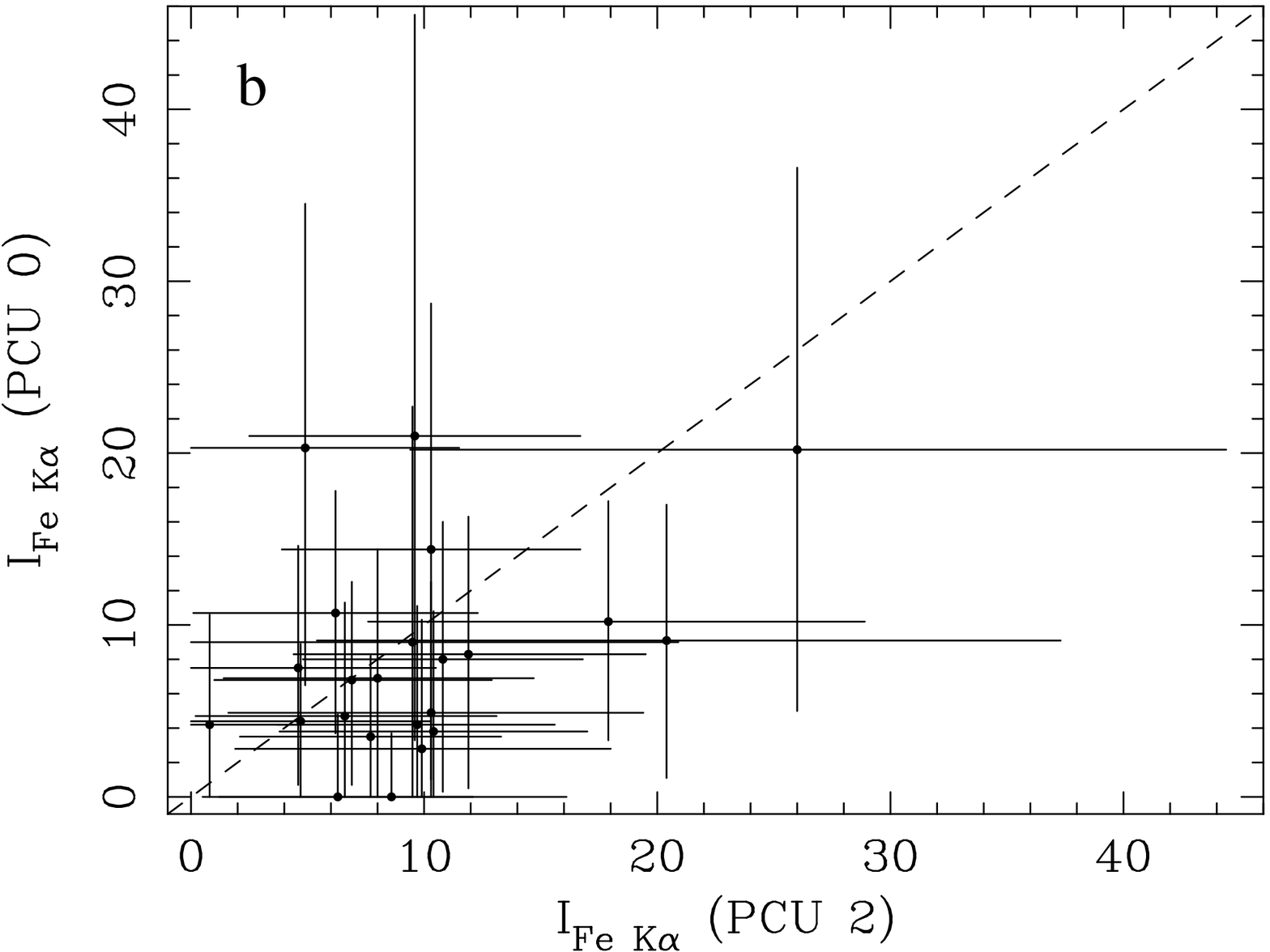}}   
 \caption{
 	(a) PCU 0 vs PCU 2 measurements for the photon index, $\Gamma$, obtained from the power-law
	plus Gaussian fits (\S\ref{indfits}).  The dotted line represents $\Gamma$(PCU 0)=$\Gamma$(PCU
	2). (b) PCU 0 vs. PCU 2 measurements for the intensity of the \feka line, obtained from the
	same fits.  The dotted line represents $I_{Fe~K\alpha}$(PCU 0)=$I_{Fe~K\alpha}$(PCU 2).
	}
\end{figure}

\begin{figure}[!htb]
 \centerline{
\epsscale{0.5}
 \scalebox{0.8}{\includegraphics[angle=270]{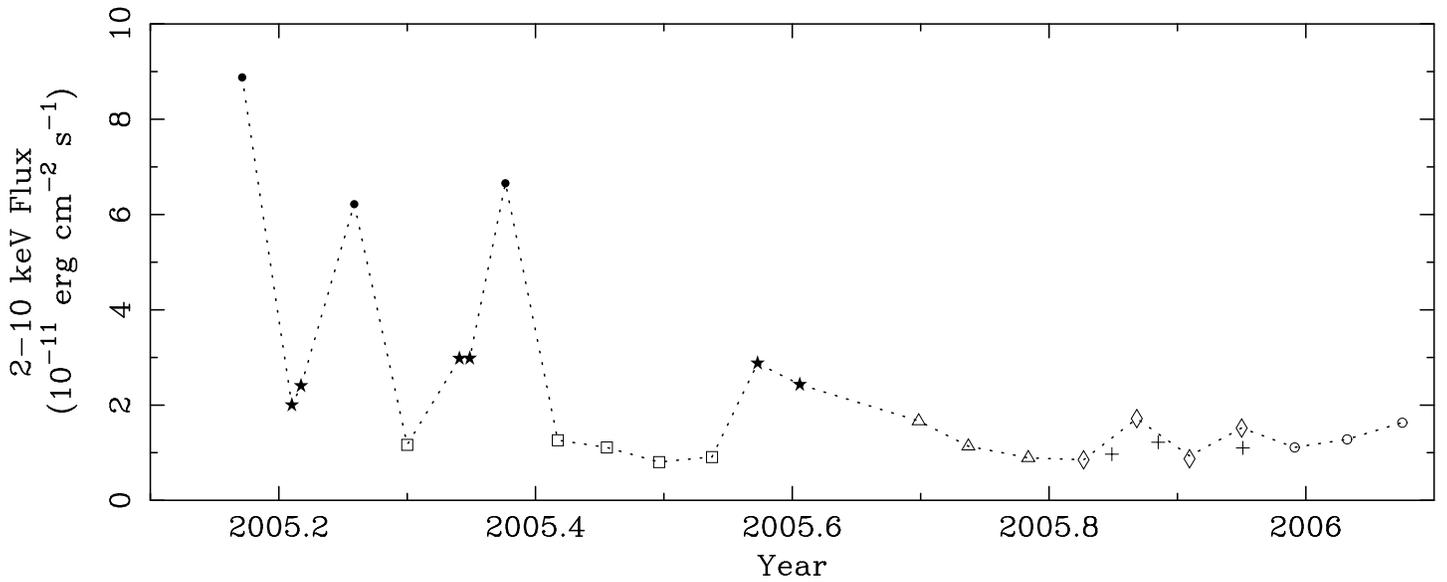}}}
 \caption{
 	The 2--10 keV light curve from the 24 RXTE observations.  Groups 1--6 are identified as follows:
	filled circles: group 1, stars: group 2, squares: group 3, triangles: group 4,
	diamonds: group 5, open circles: group 6.  Crosses represent quasi-simultaneous
	Suzaku observations.
	}
\end{figure}

\begin{figure}[!htb]
 \centerline{
   \scalebox{0.8}{\includegraphics[angle=270]{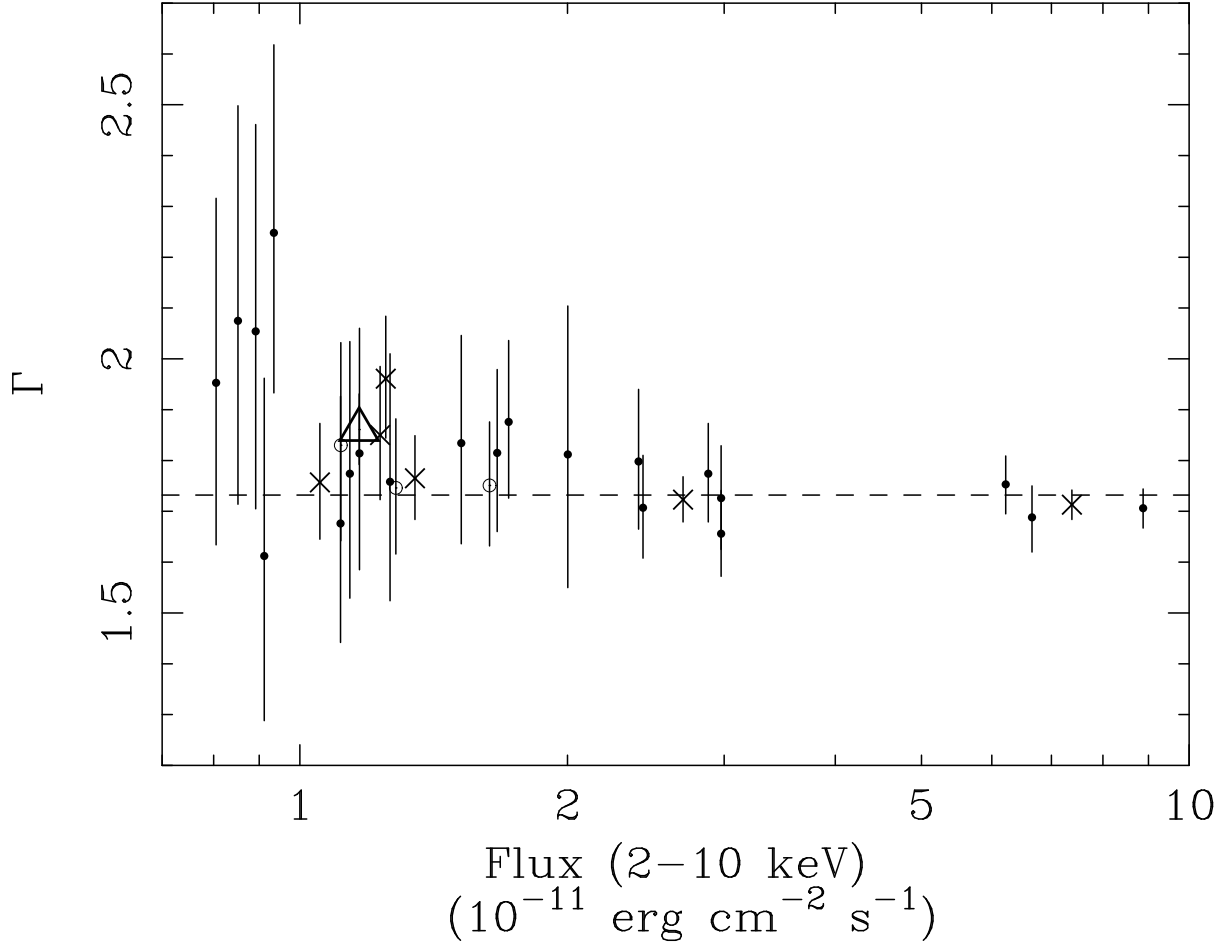}}}
 \caption{
 	Plot of $\Gamma$ vs. 2--10 keV flux for the combined PCU 0~+~PCU 2 data (open and filled
	circles).  Open circles identify those observations in which the centroid energy and
	the width of the Fe~K line had to be fixed due to weak line detection (see \tableobslogp).
	Measurements for groups 1--6 are given by crosses. 
	The measurement for group 3+4+5 is given by an open triangle.  
	The dotted line corresponds to the weighted
	mean value of $\Gamma=1.732$.
	}
\end{figure}

\begin{figure}[!htb]
 \centerline{
   \scalebox{0.8}{\includegraphics[angle=270]{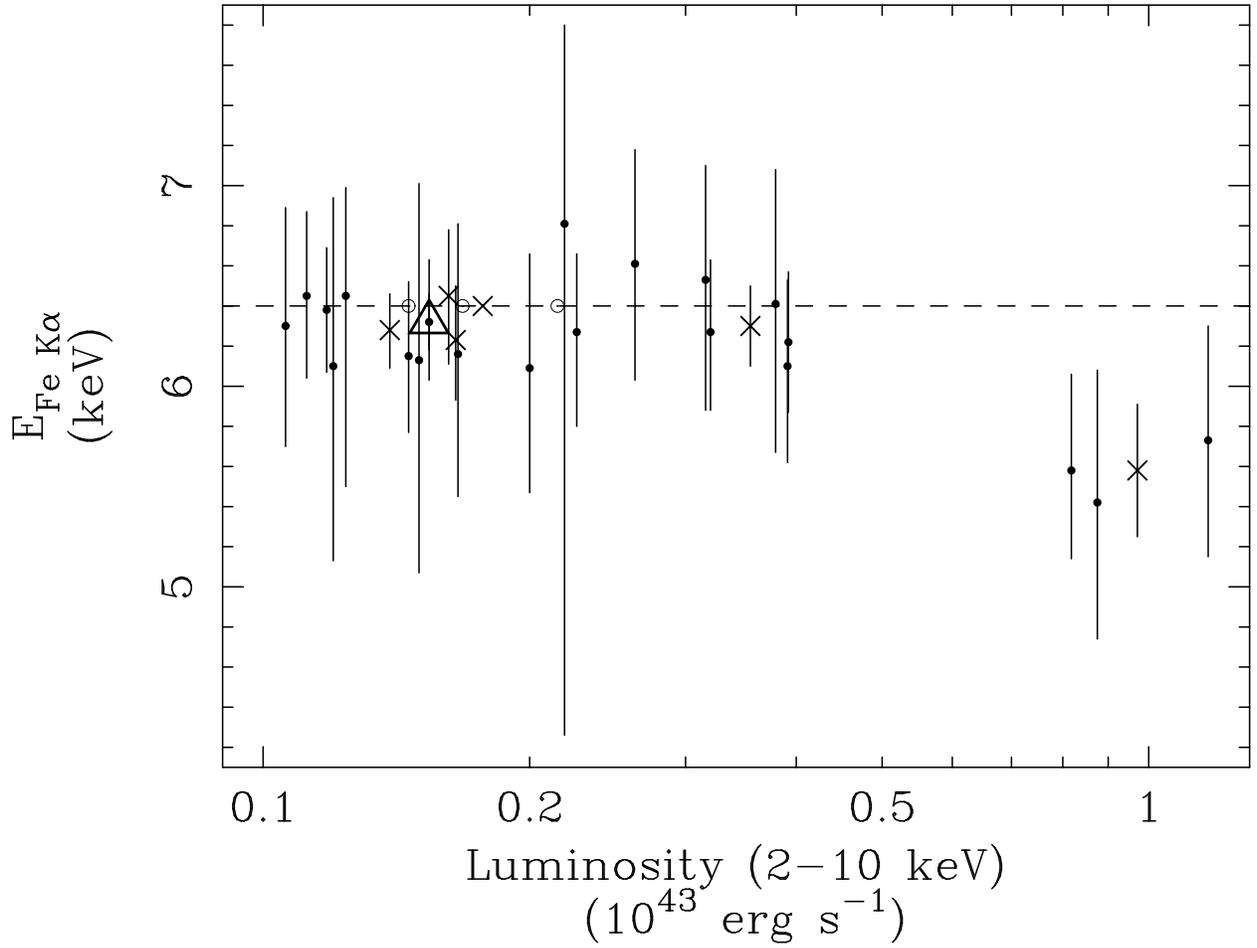}}}
 \caption{
 	Centroid energy of the \feka line vs. 2--10 keV luminosity.  See Fig. 4 caption for
	explanation of symbols.  The dotted line corresponds to the rest energy of the Fe~K
	line from neutral Fe,
	6.4 keV.
	}
\end{figure}

\begin{figure}[!htb]
 \centerline{
\epsscale{0.8}
   \plotone{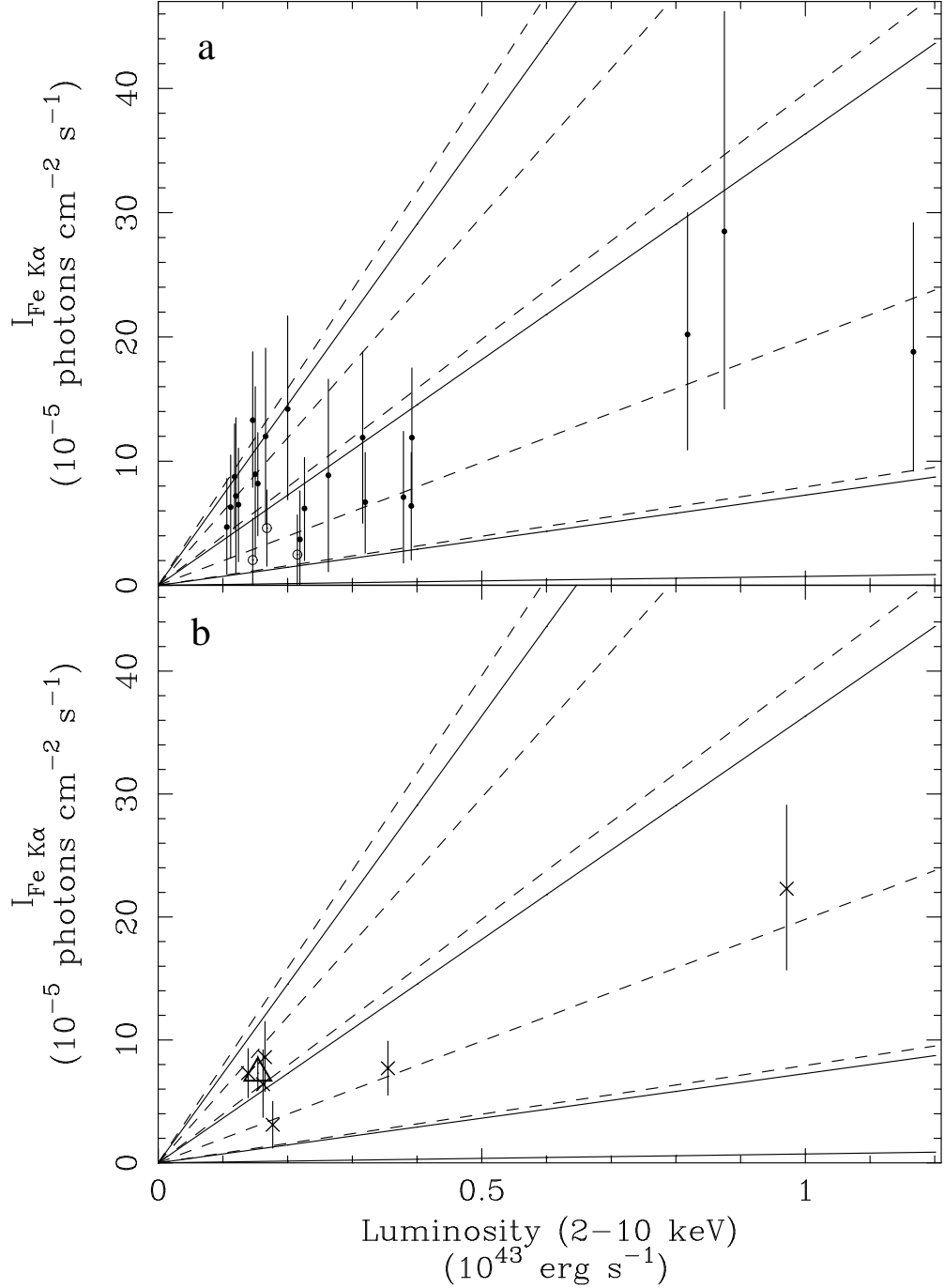}}
 \caption{
 	Intensity of the \feka line vs. 2--10 keV luminosity.  (a) Combined PCU~0~+~PCU~2 
	measurements.  Open circles identify those observations in which the centroid energy and
	the width of the Fe~K line had to be fixed due to weak line detection (see \tableobslogp).
	(b) As above, for the grouped data (crosses). 
	The measurement for group 3+4+5 is represented by an open triangle.  
	(a, b) Theoretical curves for the Compton-thin case with
	$\left(\frac{<L>}{L}\right)\left(\frac{N_{\rm H}}{10^{23} \ \rm cm^{-2}}\right)=
	$ 0.1, 1, 5, and 10 (solid lines) and
	for the Compton-thick case with $\left(\frac{<L>}{L}\right)$EW = 100,
	250, 500, 750, 
	and 1000 eV (dashed lines) are overlaid (see \S\ref{cont}).
	}
\end{figure}

\begin{figure}[!htb]
 \centerline{
   \scalebox{0.8}{\includegraphics[angle=270]{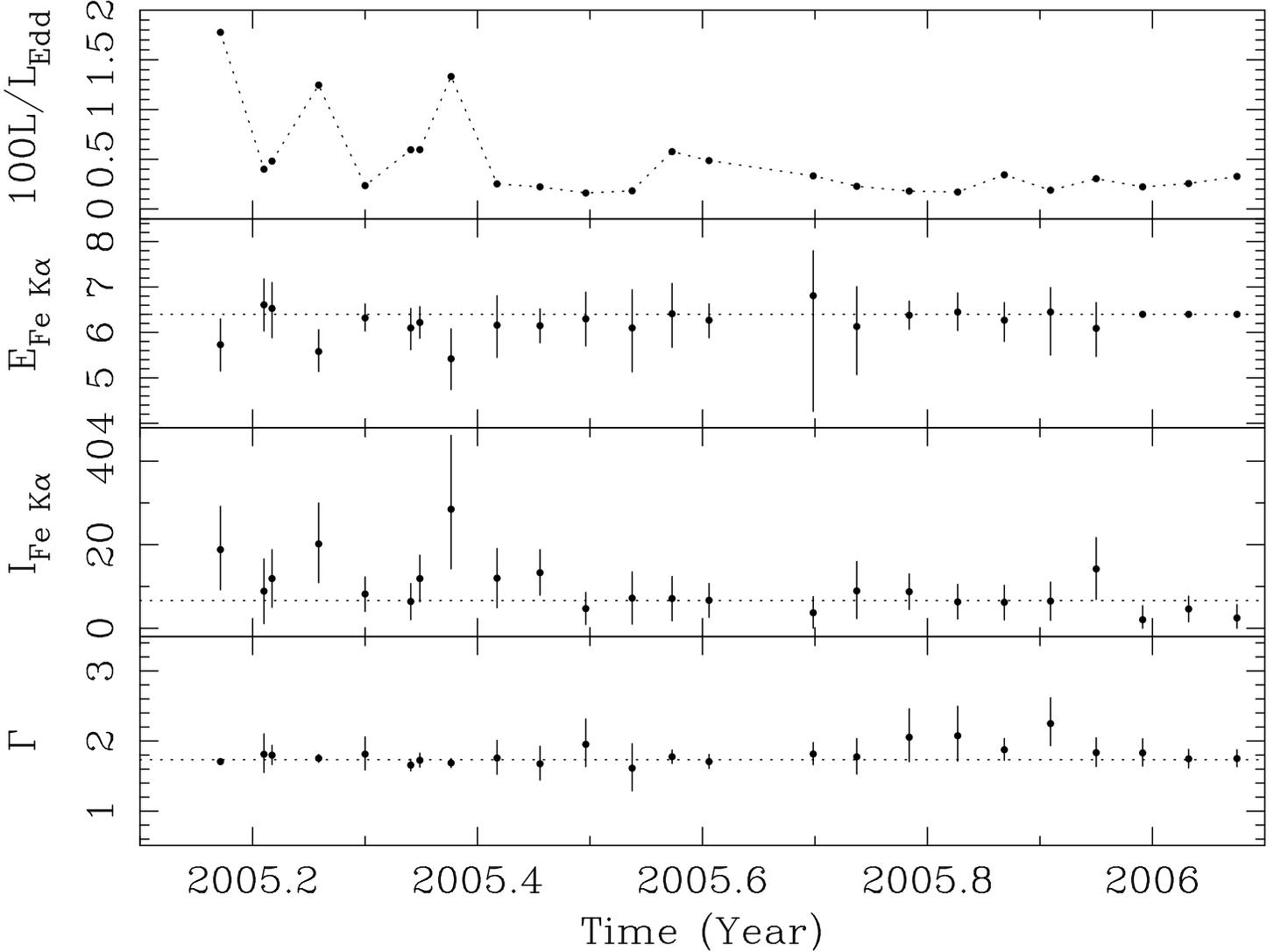}}}
 \caption{
 	Results of PCU~0~+~PCU~2 model fits vs. time (see \S\ref{indfits}).  For the fraction of the
	Eddington luminosity ($L/L_{Edd}$), a black hole mass of $5.2\times10^{7}
	M_{\odot}$ was assumed.  Also shown are the energy of the \feka line in keV, the intensity of
	the \feka line in $10^{-5} \ \rm photons \ cm^{-2} \ s^{-1}$, and the photon index. 
	The energy of the \feka line was fixed at 6.4 keV for the last three observations 
	(22, 23, \& 24) since the line was not significantly detected in these data sets.  The
	dotted lines refer to 6.4 keV in the \efeka plot and the weighted mean values of \ifeka
	and $\Gamma$ in their respective plots.
	}
\end{figure}

\begin{figure}[!htb]
 \centerline{
   \scalebox{0.8}{\includegraphics[angle=270]{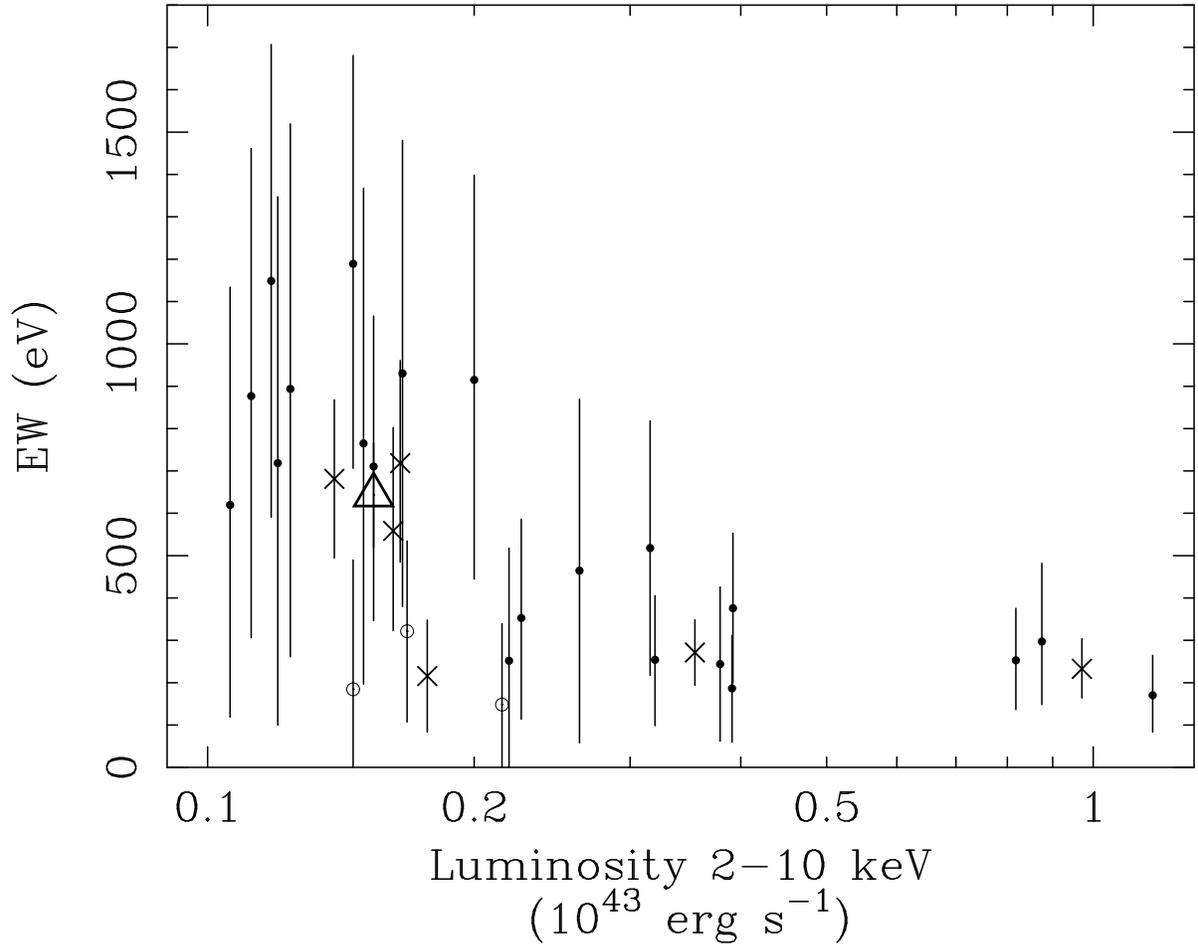}}}
 \caption{
 	Equivalent width of the \feka line vs. 2--10 keV luminosity.  See Fig. 4 caption for
	explanation of symbols.
	}
\end{figure}

\begin{figure}[!htb]
 \centerline{
\epsscale{1}
   \plotone{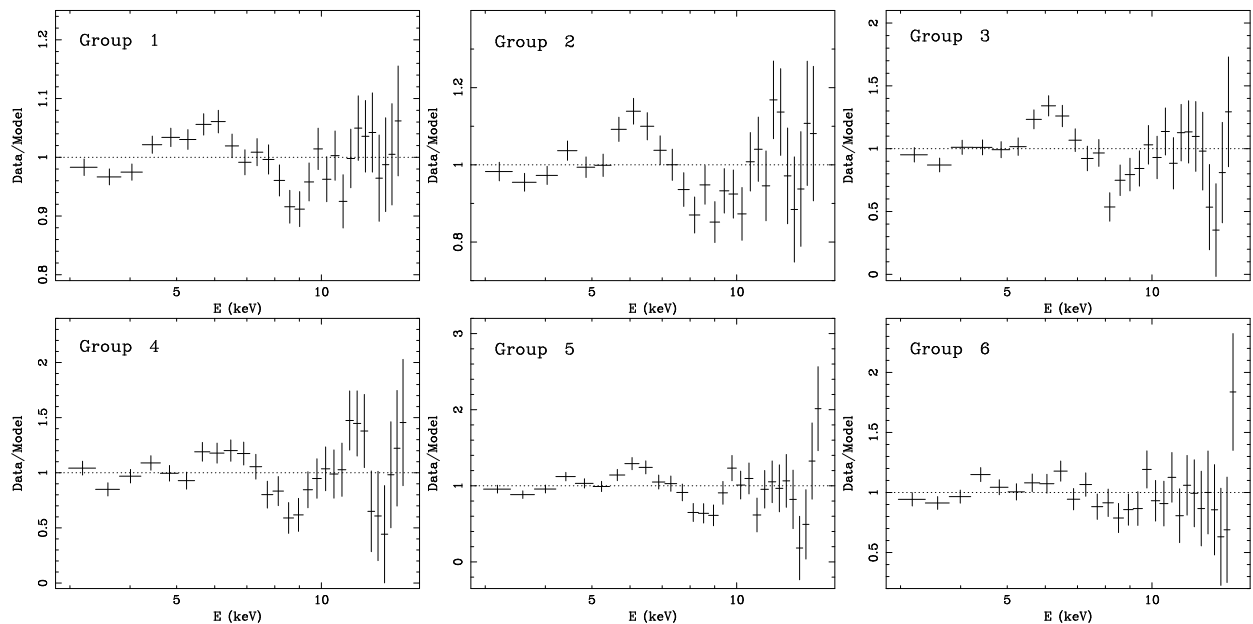}}
 \caption{
 	Plots of the ratios of the grouped data to a simple model.  The model includes
	a power-law continuum and Galactic absorption only.  The y-axis scales are inconsistent
	in order to clearly display the spectral features.
	}
\end{figure}

\begin{figure}[!htb]
 \centerline{
\epsscale{1}
   \plotone{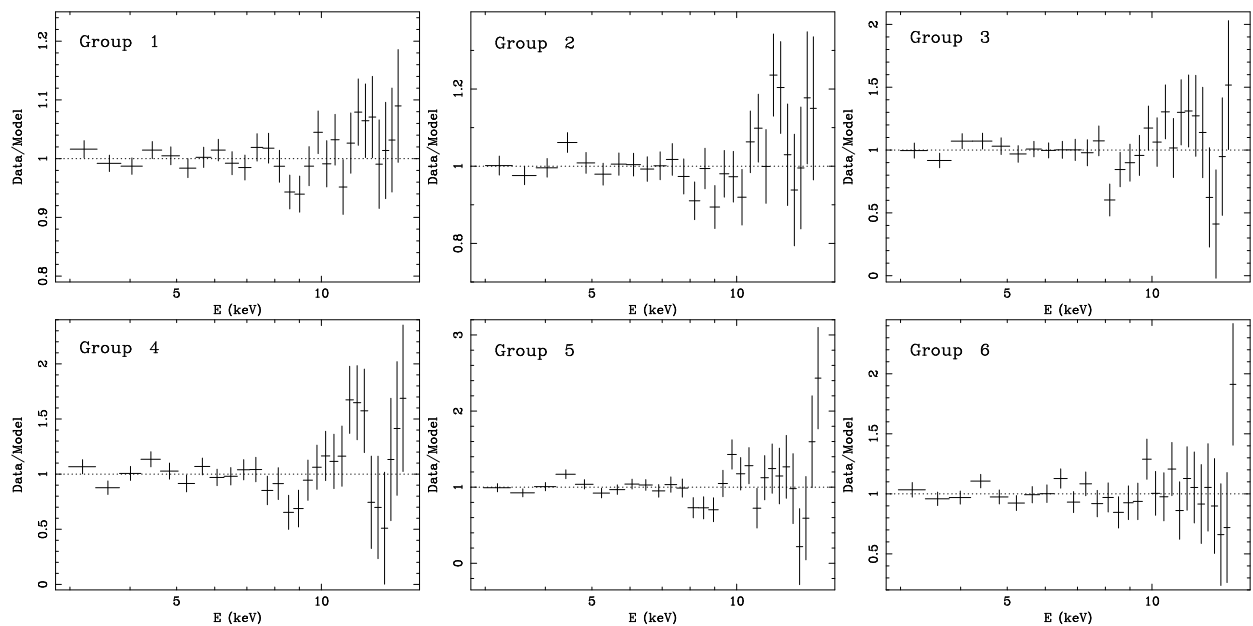}}
 \caption{
 	Plots of the ratios of the grouped data a model including \feka line emission.  
	The model includes a power-law continuum and Galactic absorption plus a 
	Gaussian component to model the \feka line.
	}
\end{figure}

\begin{figure}[!htb]
  \centerline{
    \plotone{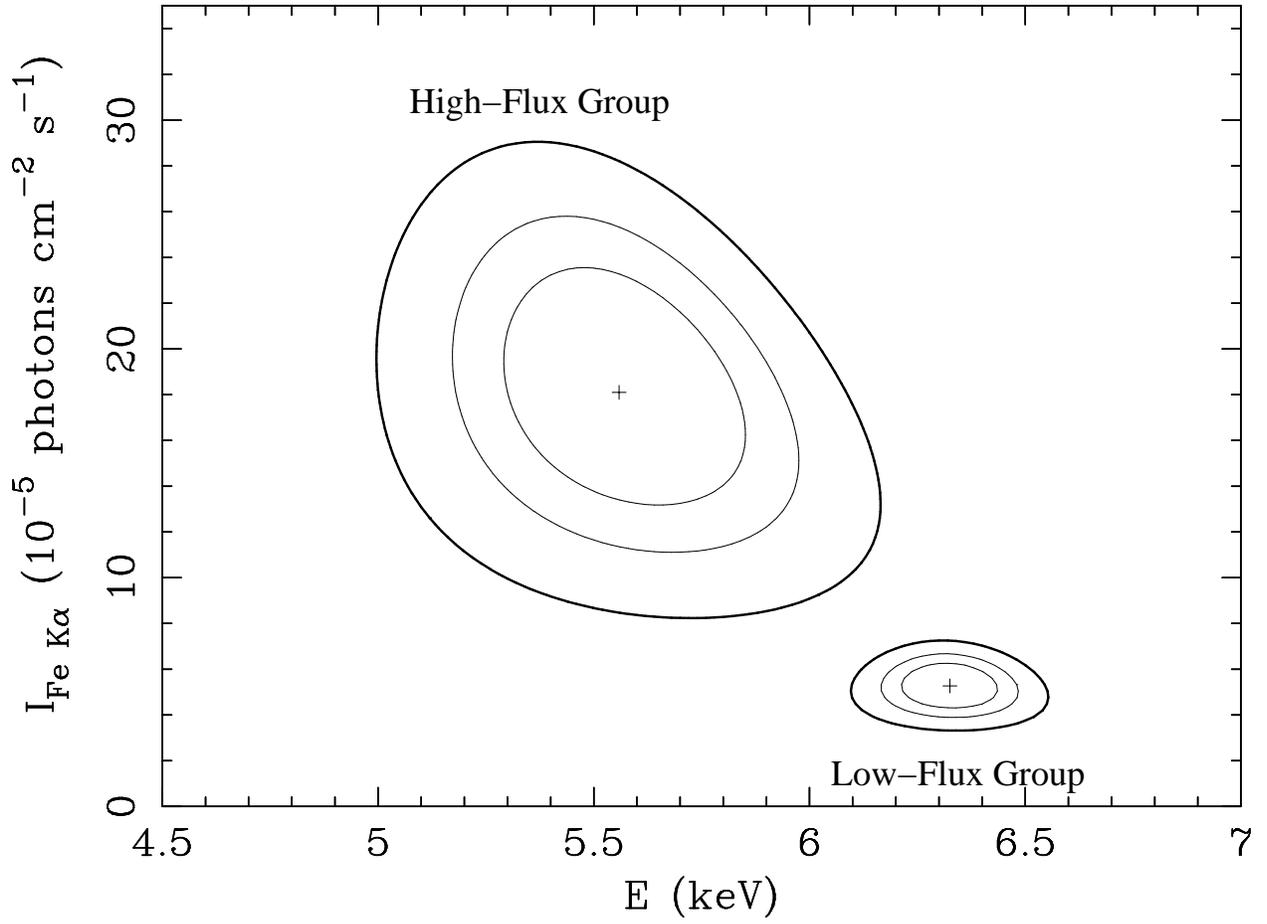}}
  \caption{
  	The 68\%, 90\%, \& 99\% confidence contours of the intensity of the \feka line versus
	its centroid energy for the high-flux group and the low-flux
	group (\S\ref{highlow}).
	}
\end{figure} 

\begin{figure}[!htb]
  \centerline{
\epsscale{1}
    \plotone{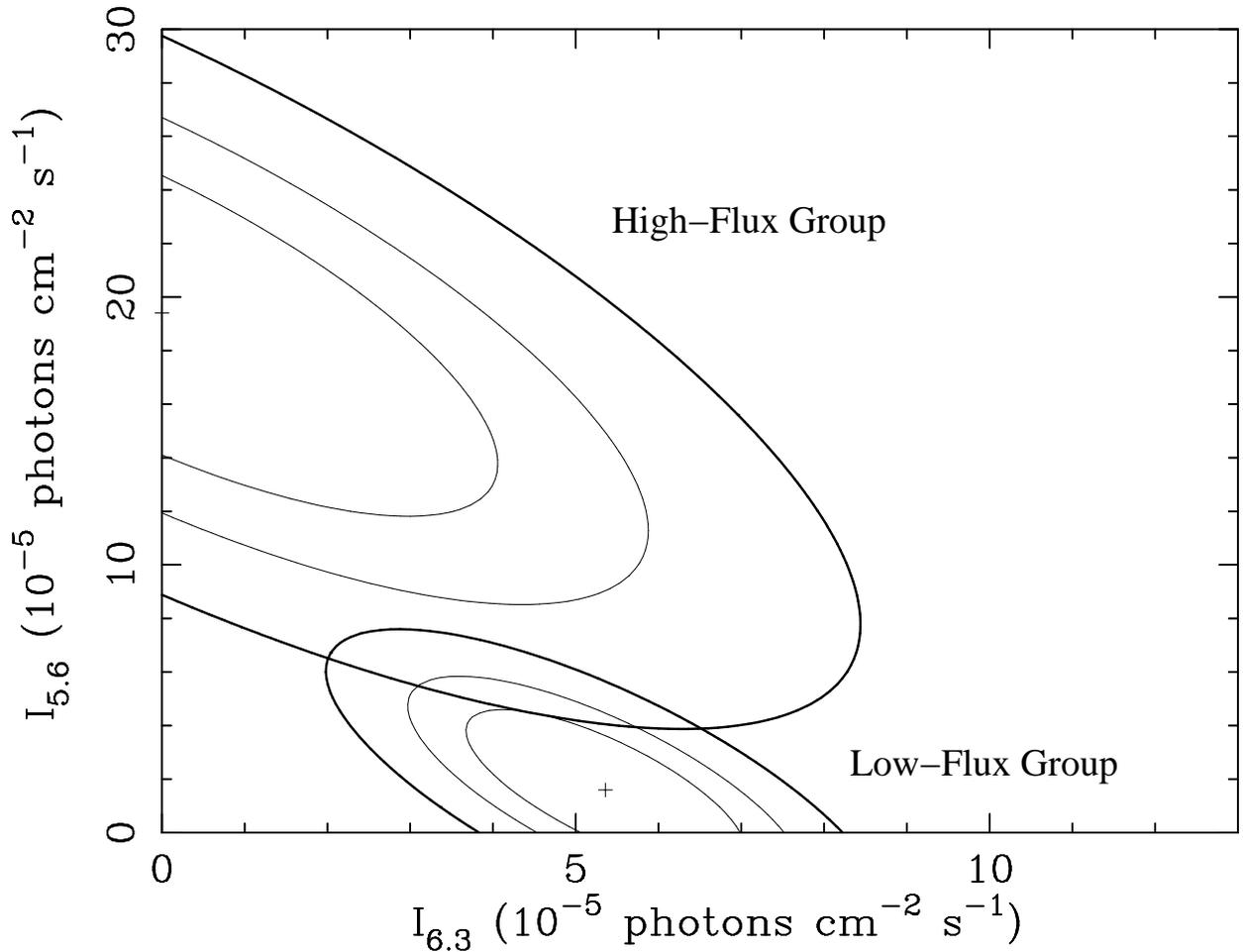}}
  \caption{
  	Constraints on a dual Gaussian-line model, fitted to the high- and low-flux data. 
	Shown are confidence contour plots (68\%, 90\%, and 99\%) of the intensity of a line 
	with the centroid energy fixed at the best-fitting value found in the 
	high-flux state ($E=5.6$ keV) versus the intensity of a line with the
	centroid energy fixed at the best-fitting value found in the 
	low-flux state ($E=6.3$ keV) for the high- and low-flux groups.
	}
\end{figure}

\begin{figure}[!htb]
\centerline{
	\scalebox{0.8}{\includegraphics[angle=270]{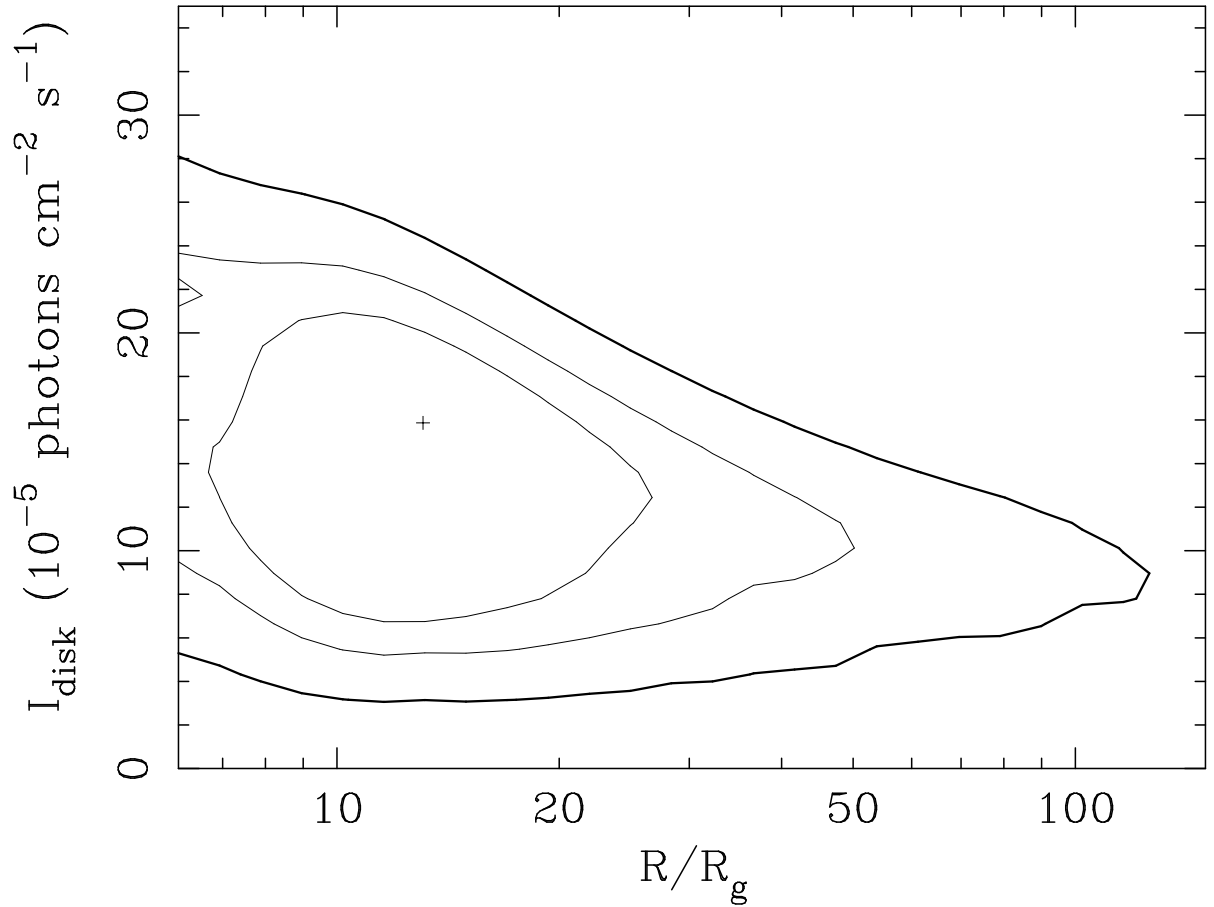}}}
	\caption{
		Constraints from the redshifted Fe~K line in the high-flux (group~1) spectrum.
		Shown are the 68\%, 90\%, \& 99\% confidence contours of the intensity of a
		relativistic Fe~K 
		disk line component versus the effective outer radius of the disk line emission in 
		units of gravitational radii.
	       }
\end{figure}

\begin{figure}[!htb]
  \centerline{
\epsscale{1}
    \plotone{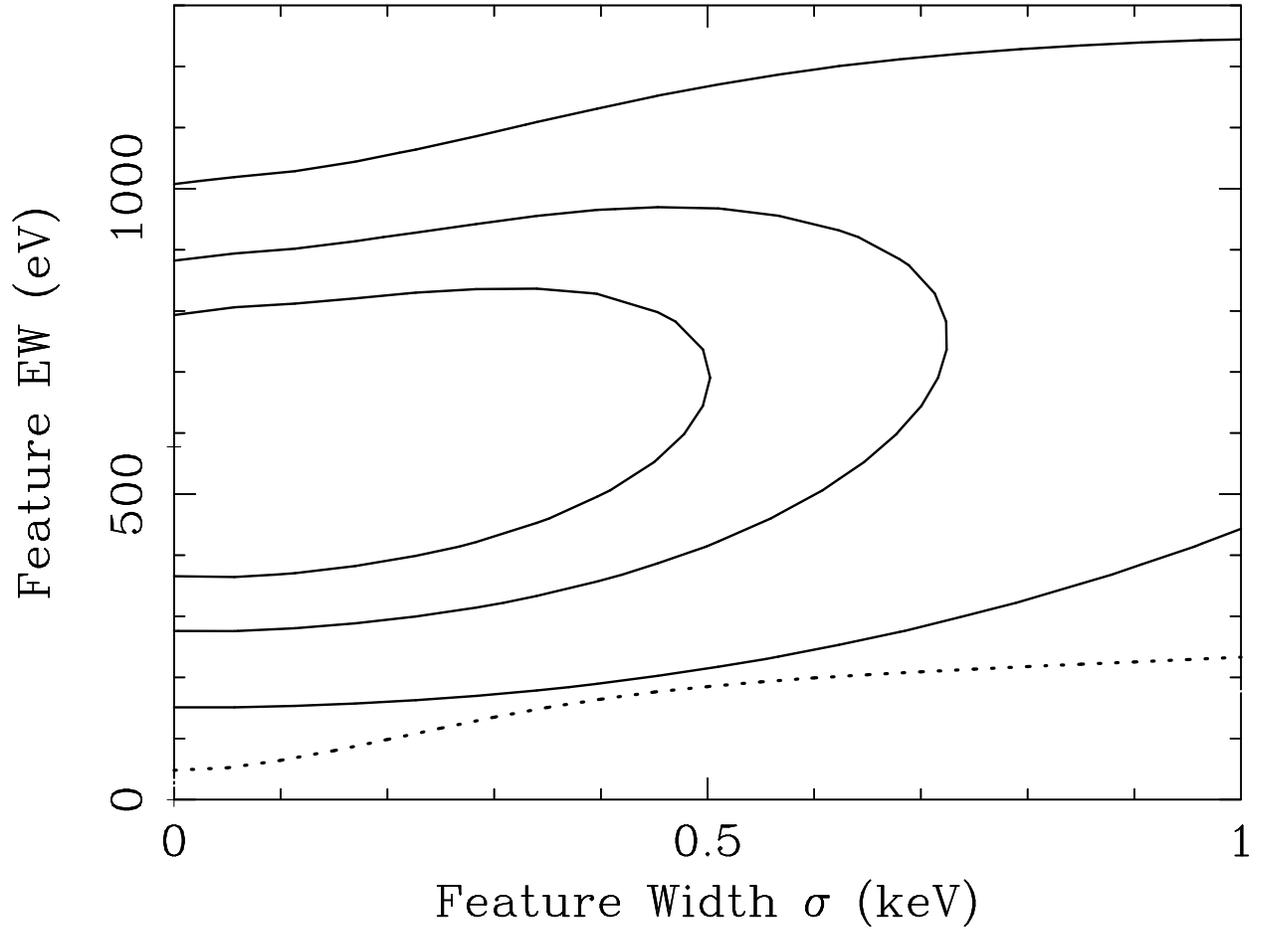}}
  \caption{
  	Confidence contours (68\%, 90\%, \& 99\%) of EW versus intrinsic width of the 
	8--9 keV feature for group 5 of the \rxte data for NGC~2992.  
	The 99\% confidence, upper limit contour of the feature for
	quasi-simultaneous Suzaku data is shown as dotted line.
	}
\end{figure}

\newpage

%Table 1
\begin{deluxetable}{ccccccc} 
\small
\tablecolumns{4} 
\tablewidth{0pc} 
\tablecaption{NGC 2992 {\it RXTE} Observation Log} 
\tablehead{ 
\colhead{Obs} & \colhead{Start\tablenotemark{a}} & & \colhead{End\tablenotemark{a}} & & \colhead{Exposure\tablenotemark{b}} & \colhead{Count Rate\tablenotemark{c}}}
\startdata
1 & 04/03/05 & 13:27:23 & 04/03/05 & 14:20:27 & 6368 & 10.590 $\pm$ 0.064\\	  
2 & 18/03/05 & 17:18:35 & 18/03/05 & 17:31:07 & 1504 & 2.287 $\pm$ 0.106\\	  
3 & 21/03/05 & 08:16:11 & 21/03/05 & 08:54:51 & 4640 & 2.730 $\pm$ 0.062\\	  
4 & 05/04/05 & 10:07:39 & 05/04/05 & 10:57:47 & 4928 & 7.467 $\pm$ 0.068\\	  
5 & 20/04/05 & 12:12:27 & 20/04/05 & 12:57:47 & 5440 & 1.326 $\pm$ 0.055\\	  
6 & 05/05/05 & 07:37:31 & 05/05/05 & 08:28:43 & 6144 & 3.646 $\pm$ 0.055\\	  
7 & 08/05/05 & 06:25:15 & 08/05/05 & 07:04:59 & 4576 & 3.673 $\pm$ 0.063\\	  
8 & 18/05/05 & 10:11:15 & 18/05/05 & 11:14:43 & 5216 & 8.084 $\pm$ 0.068\\	  
9 & 02/06/05 & 07:32:11 & 02/06/05 & 08:16:11 & 5280 & 1.555 $\pm$ 0.055\\	  
10 & 16/06/05 & 06:30:35 & 16/06/05 & 07:24:27 & 6464 & 1.350 $\pm$ 0.051\\	  
11 & 01/07/05 & 03:34:35 & 01/07/05 & 04:27:55 & 6400 & 0.802 $\pm$ 0.050\\	  
12 & 16/07/05 & 03:57:31 & 16/07/05 & 04:35:07 & 4512 & 1.203 $\pm$ 0.060\\	  
13 & 29/07/05 & 03:19:39 & 29/07/05 & 04:10:03 & 6048 & 3.402 $\pm$ 0.056\\	  
14 & 10/08/05 & 03:07:55 & 10/08/05 & 04:01:31 & 6432 & 2.841 $\pm$ 0.053\\	  
15 & 12/09/05 & 22:44:27 & 12/09/05 & 23:35:07 & 5600 & 1.882 $\pm$ 0.053\\	  
16 & 27/09/05 & 00:55:55 & 27/09/05 & 01:46:51 & 6112 & 1.322 $\pm$ 0.051\\	  
17 & 14/10/05 & 01:45:31 & 14/10/05 & 02:39:07 & 5600 & 0.886 $\pm$ 0.055\\	  
18 & 29/10/05 & 19:13:31 & 29/10/05 & 20:03:07 & 5632 & 1.013 $\pm$ 0.054\\	  
19 & 13/11/05 & 22:35:23 & 13/11/05 & 23:23:07 & 5728 & 1.970 $\pm$ 0.054\\	  
20 & 28/11/05 & 22:34:03 & 28/11/05 & 23:27:23 & 6400 & 1.687 $\pm$ 0.051\\	  
21 & 13/12/05 & 17:55:07 & 13/12/05 & 18:45:31 & 6048 & 1.827 $\pm$ 0.055\\	  
22 & 28/12/05 & 19:46:51 & 28/12/05 & 20:30:35 & 5248 & 1.278 $\pm$ 0.057\\	  
23 & 12/01/06 & 18:06:34 & 12/01/06 & 18:59:54 & 6400 & 0.850 $\pm$ 0.051\\	  
24 & 28/01/06 & 09:52:11 & 28/01/06 & 10:45:31 & 6368 & 1.882 $\pm$ 0.053\\	  
\enddata 
\tablenotetext{a}{Universal time} 
\tablenotetext{b}{Exposure time is given in seconds} 
\tablenotetext{c}{Count rate is given in counts s$^{-1}$}
\end{deluxetable} 

%Table 2
\begin{deluxetable}{rccccccccc}
\rotate
\tablecolumns{15}
\tablewidth{0pc}
\tablecaption{NGC 2992 {\it RXTE} Fit Results}
\tablehead{ 
\colhead{Obs} & \colhead{F$_{2-10}$\tablenotemark{a}} & \colhead{L$_{2-10}$\tablenotemark{b}} & \colhead{$\Gamma$\tablenotemark{c}} & \colhead{\efeka\tablenotemark{d}} & \colhead{\sigfeka\tablenotemark{e}} & \colhead{\ifeka\tablenotemark{f}} & \colhead{EW\tablenotemark{g}} & \colhead{$\chi^{2}$\tablenotemark{h}} & \colhead{dof\tablenotemark{i}}}
\startdata
 1 &  8.88 &  1.17 & $ 1.706 ^{+0.038} _{-0.039} $ & $  5.73 ^{+ 0.57} _{- 0.58} $ & $  0.83 ^{+  0.55} _{- 0.49} $ & $  18.8 ^{+ 10.4} _{-  9.6} $ & $ 170^{+  94} _{-  86} $ & 20.1 & 23 \\
 & & & [1.672-1.739] & [ 5.22- 6.23] & [ 0.41-  1.30] & [  10.3-  27.8] & [  93- 251] & & \\
 2 &  2.00 &  0.26 & $ 1.812 ^{+0.292} _{-0.262} $ & $  6.61 ^{+ 0.57} _{- 0.58} $ & $  0.00 ^{+  1.03} _{- 0.00} $ & $   8.9 ^{+  7.7} _{-  7.8} $ & $ 464^{+ 405} _{- 406} $ & 17.5 & 23 \\
 & & & [1.581-2.066] & [ 6.11- 7.00] & [ 0.00-  0.81] & [   1.8-  15.6] & [  94- 817] & & \\
 3 &  2.40 &  0.32 & $ 1.798 ^{+0.142} _{-0.133} $ & $  6.53 ^{+ 0.57} _{- 0.65} $ & $  0.68 ^{+  0.80} _{- 0.68} $ & $  11.9 ^{+  6.9} _{-  6.9} $ & $ 517^{+ 300} _{- 300} $ & 10.6 & 23 \\
 & & & [1.681-1.922] & [ 5.98- 7.02] & [ 0.00-  1.35] & [   5.9-  18.0] & [ 256- 783] & & \\
 4 &  6.22 &  0.82 & $ 1.753 ^{+0.056} _{-0.058} $ & $  5.58 ^{+ 0.48} _{- 0.44} $ & $  0.68 ^{+  0.71} _{- 0.68} $ & $  20.2 ^{+  9.8} _{-  9.3} $ & $ 252^{+ 122} _{- 116} $ & 17.1 & 23 \\
 & & & [1.703-1.803] & [ 5.19- 5.99] & [ 0.00-  1.30] & [  12.0-  28.8] & [ 150- 360] & & \\
 5 &  1.17 &  0.15 & $ 1.814 ^{+0.246} _{-0.229} $ & $  6.32 ^{+ 0.31} _{- 0.29} $ & $  0.00 ^{+  0.51} _{- 0.00} $ & $   8.2 ^{+  4.1} _{-  4.2} $ & $ 710^{+ 355} _{- 363} $ & 24.1 & 23 \\
 & & & [1.612-2.029] & [ 6.11- 6.54] & [ 0.00-  0.43] & [   4.5-  11.8] & [ 389-1022] & & \\
 6 &  2.98 &  0.39 & $ 1.656 ^{+0.085} _{-0.084} $ & $  6.10 ^{+ 0.43} _{- 0.48} $ & $  0.04 ^{+  1.09} _{- 0.04} $ & $   6.4 ^{+  4.3} _{-  4.4} $ & $ 186^{+ 125} _{- 126} $ & 22.1 & 23 \\
 & & & [1.582-1.731] & [ 5.70- 6.46] & [ 0.00-  0.98] & [   2.6-  10.2] & [  75- 297] & & \\
 7 &  2.98 &  0.39 & $ 1.726 ^{+0.103} _{-0.101} $ & $  6.22 ^{+ 0.35} _{- 0.35} $ & $  0.33 ^{+  0.57} _{- 0.33} $ & $  11.9 ^{+  5.6} _{-  5.6} $ & $ 375^{+ 176} _{- 176} $ & 20.3 & 23 \\
 & & & [1.637-1.816] & [ 5.92- 6.53] & [ 0.00-  0.80] & [   7.0-  16.8] & [ 221- 530] & & \\
 8 &  6.66 &  0.88 & $ 1.688 ^{+0.062} _{-0.068} $ & $  5.42 ^{+ 0.66} _{- 0.68} $ & $  1.10 ^{+  0.70} _{- 0.64} $ & $  28.5 ^{+ 17.7} _{- 14.3} $ & $ 297^{+ 184} _{- 149} $ & 14.3 & 23 \\
 & & & [1.629-1.743] & [ 4.82- 5.99] & [ 0.53-  1.71] & [  15.8-  43.8] & [ 164- 456] & & \\
 9 &  1.26 &  0.17 & $ 1.758 ^{+0.252} _{-0.234} $ & $  6.16 ^{+ 0.65} _{- 0.71} $ & $  0.80 ^{+  0.89} _{- 0.80} $ & $  12.0 ^{+  7.1} _{-  7.1} $ & $ 930^{+ 550} _{- 550} $ & 13.2 & 23 \\
 & & & [1.554-1.978] & [ 5.55- 6.72] & [ 0.00-  1.54] & [   5.8-  18.2] & [ 449-1410] & & \\
10 &  1.11 &  0.15 & $ 1.676 ^{+0.250} _{-0.234} $ & $  6.15 ^{+ 0.37} _{- 0.38} $ & $  0.59 ^{+  0.53} _{- 0.59} $ & $  13.3 ^{+  5.5} _{-  5.4} $ & $1189^{+ 491} _{- 482} $ & 15.7 & 23 \\
 & & & [1.470-1.894] & [ 5.82- 6.47] & [ 0.00-  1.04] & [   8.6-  18.1] & [ 768-1618] & & \\

\tablebreak

11 &  0.81 &  0.11 & $ 1.953 ^{+0.363} _{-0.319} $ & $  6.30 ^{+ 0.59} _{- 0.60} $ & $  0.00 ^{+  1.19} _{- 0.00} $ & $   4.7 ^{+  3.9} _{-  3.8} $ & $ 619^{+ 514} _{- 501} $ &  6.0 & 23 \\
 & & & [1.670-2.269] & [ 5.80- 6.79] & [ 0.00-  0.98] & [   1.4-   8.1] & [ 178-1068] & & \\
12 &  0.91 &  0.12 & $ 1.612 ^{+0.350} _{-0.324} $ & $  6.10 ^{+ 0.84} _{- 0.97} $ & $  0.55 ^{+  1.01} _{- 0.55} $ & $   7.2 ^{+  6.3} _{-  6.2} $ & $ 718^{+ 628} _{- 618} $ & 15.6 & 23 \\
 & & & [1.328-1.916] & [ 5.30- 6.80] & [ 0.00-  1.35] & [   1.8-  12.7] & [ 179-1267] & & \\
13 &  2.88 &  0.38 & $ 1.774 ^{+0.099} _{-0.095} $ & $  6.41 ^{+ 0.67} _{- 0.74} $ & $  0.46 ^{+  1.19} _{- 0.46} $ & $   7.1 ^{+  5.3} _{-  5.3} $ & $ 243^{+ 182} _{- 182} $ & 14.8 & 23 \\
 & & & [1.691-1.861] & [ 5.80- 6.98] & [ 0.00-  1.45] & [   2.5-  11.8] & [  85- 405] & & \\
14 &  2.43 &  0.32 & $ 1.707 ^{+0.103} _{-0.099} $ & $  6.27 ^{+ 0.36} _{- 0.39} $ & $  0.00 ^{+  1.02} _{- 0.00} $ & $   6.7 ^{+  4.0} _{-  4.1} $ & $ 253^{+ 151} _{- 155} $ & 17.9 & 23 \\
 & & & [1.620-1.797] & [ 5.95- 6.57] & [ 0.00-  0.89] & [   3.1-  10.2] & [ 117- 386] & & \\
15 &  1.67 &  0.22 & $ 1.815 ^{+0.164} _{-0.155} $ & $  6.81 ^{+ 0.99} _{- 2.55} $ & $  0.00 ^{+  1.94} _{- 0.00} $ & $   3.7 ^{+  3.9} _{-  3.7} $ & $ 251^{+ 265} _{- 251} $ & 17.2 & 24 \\
 & & & [1.679-1.958] & [ 6.08- 7.42] & [ 0.00-  1.09] & [   0.4-   7.1] & [  27- 483] & & \\
16 &  1.14 &  0.15 & $ 1.774 ^{+0.260} _{-0.245} $ & $  6.13 ^{+ 0.88} _{- 1.06} $ & $  0.86 ^{+  1.04} _{- 0.86} $ & $   8.9 ^{+  7.1} _{-  6.6} $ & $ 764^{+ 602} _{- 568} $ & 21.4 & 24 \\
 & & & [1.559-2.001] & [ 5.24- 6.89] & [ 0.00-  1.70] & [   3.0-  15.1] & [ 256-1290] & & \\
17 &  0.89 &  0.12 & $ 2.054 ^{+0.407} _{-0.349} $ & $  6.38 ^{+ 0.31} _{- 0.31} $ & $  0.07 ^{+  0.70} _{- 0.07} $ & $   8.8 ^{+  4.2} _{-  4.2} $ & $1148^{+ 557} _{- 557} $ & 22.7 & 24 \\
 & & & [1.745-2.407] & [ 6.11- 6.65] & [ 0.00-  0.69] & [   5.0-  12.5] & [ 656-1641] & & \\
18 &  0.85 &  0.11 & $ 2.075 ^{+0.423} _{-0.361} $ & $  6.45 ^{+ 0.42} _{- 0.41} $ & $  0.03 ^{+  0.75} _{- 0.03} $ & $   6.3 ^{+  4.2} _{-  4.1} $ & $ 876^{+ 584} _{- 570} $ & 32.5 & 24 \\
 & & & [1.754-2.442] & [ 6.10- 6.81] & [ 0.00-  0.68] & [   2.7-  10.0] & [ 375-1391] & & \\
19 &  1.72 &  0.23 & $ 1.876 ^{+0.160} _{-0.150} $ & $  6.27 ^{+ 0.39} _{- 0.47} $ & $  0.03 ^{+  1.42} _{- 0.03} $ & $   6.2 ^{+  4.1} _{-  4.2} $ & $ 352^{+ 233} _{- 238} $ & 13.5 & 24 \\
 & & & [1.744-2.016] & [ 5.88- 6.62] & [ 0.00-  1.22] & [   2.5-   9.8] & [ 142- 557] & & \\
20 &  0.94 &  0.12 & $ 2.248 ^{+0.370} _{-0.315} $ & $  6.45 ^{+ 0.54} _{- 0.95} $ & $  0.40 ^{+  0.89} _{- 0.40} $ & $   6.5 ^{+  4.6} _{-  4.6} $ & $ 893^{+ 625} _{- 632} $ & 27.9 & 24 \\
 & & & [1.969-2.570] & [ 5.79- 6.92] & [ 0.00-  1.14] & [   2.5-  10.5] & [ 343-1443] & & \\

\tablebreak

21 &  1.52 &  0.20 & $ 1.834 ^{+0.212} _{-0.198} $ & $  6.09 ^{+ 0.57} _{- 0.62} $ & $  0.84 ^{+  0.71} _{- 0.84} $ & $  14.2 ^{+  7.5} _{-  7.3} $ & $ 915^{+ 483} _{- 470} $ & 14.1 & 24 \\
 & & & [1.660-2.020] & [ 5.55- 6.59] & [ 0.00-  1.44] & [   7.8-  20.8] & [ 502-1340] & & \\
22 &  1.11 &  0.15 & $ 1.830 ^{+0.202} _{-0.188} $ & $  6.40f $ & $  0.05f $ & $   2.0 ^{+  3.4} _{-  2.0} $ & $ 184^{+ 305} _{- 184} $ & 22.8 & 26 \\
 & & & [1.626-2.052] & [\ldots] & [\ldots] & [   0.0-   5.7] & [   0- 517] & & \\
23 &  1.28 &  0.17 & $ 1.746 ^{+0.136} _{-0.130} $ & $  6.40f $ & $  0.05f $ & $   4.6 ^{+  3.1} _{-  3.1} $ & $ 321^{+ 212} _{- 214} $ & 20.7 & 26 \\
 & & & [1.605-1.894] & [\ldots] & [\ldots] & [   1.3-   8.0] & [  87- 553] & & \\
24 &  1.63 &  0.22 & $ 1.751 ^{+0.125} _{-0.119} $ & $  6.40f $ & $  0.05f $ & $   2.5 ^{+  3.2} _{-  2.5} $ & $ 148^{+ 191} _{- 148} $ & 17.2 & 26 \\
 & & & [1.622-1.887] & [\ldots] & [\ldots] & [   0.0-   6.0] & [   0- 355] & & \\
\enddata
\tablenotetext{a}{2--10 keV flux ($10^{-11}$ photons s$^{-1}$)}
\tablenotetext{b}{2--10 keV luminosity ($10^{43}$ ergs s$^{-1}$)}
\tablenotetext{c}{Photon index}
\tablenotetext{d}{Energy of the \feka line (keV)}
\tablenotetext{e}{Intrinsic width of the \feka line (keV)}
\tablenotetext{f}{Intensity of the \feka line ($10^{-5}$ photons cm$^{-2}$ s$^{-1}$)}
\tablenotetext{g}{Equivalent width of the \feka line (eV)}
\tablenotetext{h}{$\chi^{2}$ for the model}
\tablenotetext{i}{Degrees of freedom}
\end{deluxetable}

%Table 3	
\begin{deluxetable}{rcccccccccc}
\rotate
\tablecolumns{15}
\tablewidth{0pc}
\tablecaption{NGC 2992 {\it RXTE} Fit Results: Grouped Data}
\tablehead{ 
\colhead{Group} & \colhead{F$_{2-10}$\tablenotemark{a}} & \colhead{L$_{2-10}$\tablenotemark{b}} & \colhead{$\Gamma$\tablenotemark{c}} & \colhead{\efeka\tablenotemark{d}} & \colhead{\sigfeka\tablenotemark{e}} & \colhead{\ifeka\tablenotemark{f}} & \colhead{EW\tablenotemark{g}} & \colhead{$\chi^{2}$\tablenotemark{h}} & \colhead{dof\tablenotemark{i}} & \colhead{P\tablenotemark{j}}}
\startdata
 1 &  7.38 &  0.97 & $ 1.713 ^{+0.029} _{-0.029} $ & $  5.58 ^{+ 0.33} _{- 0.33} $ & $  0.89 ^{+ 0.36} _{- 0.35} $ & $ 22.3 ^{+ 6.8} _{- 6.6} $ & $ 232^{+  70} _{-  68} $ & 23.6 & 23 & 57.7 \\
 & & & [1.688-1.738] & [ 5.29- 5.87] & [ 0.58- 1.20] & [16.5- 28.3] & [ 172- 295] & & \\
 2 &  2.70 &  0.35 & $ 1.723 ^{+0.045} _{-0.044} $ & $  6.30 ^{+ 0.20} _{- 0.20} $ & $  0.32 ^{+ 0.39} _{- 0.32} $ & $  7.7 ^{+ 2.2} _{- 2.2} $ & $ 271^{+  77} _{-  77} $ & 28.2 & 23 & 79.3 \\
 & & & [1.684-1.763] & [ 6.13- 6.48] & [ 0.00- 0.66] & [ 5.7-  9.6] & [ 200- 337] & & \\
 3 &  1.05 &  0.14 & $ 1.757 ^{+0.116} _{-0.112} $ & $  6.28 ^{+ 0.18} _{- 0.19} $ & $  0.26 ^{+ 0.38} _{- 0.26} $ & $  7.3 ^{+ 2.0} _{- 2.0} $ & $ 681^{+ 186} _{- 186} $ & 27.4 & 23 & 76.2 \\
 & & & [1.658-1.859] & [ 6.12- 6.44] & [ 0.00- 0.59] & [ 5.5-  9.1] & [ 513- 849] & & \\
 4 &  1.23 &  0.16 & $ 1.850 ^{+0.135} _{-0.127} $ & $  6.45 ^{+ 0.33} _{- 0.34} $ & $  0.40 ^{+ 0.45} _{- 0.40} $ & $  6.4 ^{+ 2.8} _{- 2.7} $ & $ 558^{+ 244} _{- 235} $ & 37.4 & 24 & 96.0 \\
 & & & [1.738-1.968] & [ 6.15- 6.74] & [ 0.00- 0.79] & [ 4.0-  8.9] & [ 348- 776] & & \\
 5 &  1.25 &  0.17 & $ 1.961 ^{+0.123} _{-0.117} $ & $  6.23 ^{+ 0.27} _{- 0.30} $ & $  0.56 ^{+ 0.39} _{- 0.56} $ & $  8.6 ^{+ 2.9} _{- 2.8} $ & $ 718^{+ 242} _{- 233} $ & 42.9 & 24 & 99.0 \\
 & & & [1.858-2.069] & [ 5.96- 6.47] & [ 0.00- 0.91] & [ 6.1- 11.1] & [ 509- 927] & & \\
 6 &  1.35 &  0.18 & $ 1.765 ^{+0.084} _{-0.081} $ & $  6.40f $ & $  0.05f $ & $  3.1 ^{+ 1.9} _{- 1.9} $ & $ 215^{+ 132} _{- 132} $ & 26.8 & 26 & 58.2 \\
 & & & [1.677-1.856] & [\ldots] & [\ldots] & [ 1.0-  5.1] & [  69- 354] & & \\
3+4+5 &  1.17 &  0.15 & $ 1.861 ^{+0.070} _{-0.069} $ & $  6.32 ^{+ 0.14} _{- 0.14} $ & $  0.37 ^{+ 0.25} _{- 0.37} $ & $  7.3 ^{+ 1.4} _{- 1.4} $ & $ 642^{+ 123} _{- 123} $ & 79.4 & 23 & $>4\sigma$ \\
 & & & [1.800-1.923] & [ 6.20- 6.44] & [ 0.00- 0.59] & [ 6.0-  8.5] & [ 528- 748] & & \\
\enddata
\tablenotetext{a}{2--10 keV flux ($10^{-11}$ photons s$^{-1}$)}
\tablenotetext{b}{2--10 keV luminosity ($10^{43}$ ergs s$^{-1}$)}
\tablenotetext{c}{Photon index}
\tablenotetext{d}{Energy of the \feka line (keV)}
\tablenotetext{e}{Intrinsic width of the \feka line (keV)}
\tablenotetext{f}{Intensity of the \feka line ($10^{-5}$ photons cm$^{-2}$ s$^{-1}$)}
\tablenotetext{g}{Equivalent width of the \feka line (eV)}
\tablenotetext{h}{$\chi^{2}$ for the model}
\tablenotetext{i}{Degrees of freedom}
\tablenotetext{j}{The values of $1-\frac{P}{100}$ correspond to the probability of obtaining the
observed $\chi^{2}$ or higher from statistical fluctuations alone.}
\end{deluxetable}

\end{document}